%% file: main.tex
\documentclass[lettersize,journal]{IEEEtran}
\usepackage{amsmath,amsfonts}
\usepackage{algorithmic}
\usepackage{array}
\usepackage[caption=false,font=normalsize,labelfont=sf,textfont=sf]{subfig}
\usepackage{textcomp}
\usepackage{stfloats}
\usepackage{url}
\usepackage{verbatim}
\usepackage{graphicx}
\def\BibTeX{{\rm B\kern-.05em{\sc i\kern-.025em b}\kern-.08em
    T\kern-.1667em\lower.7ex\hbox{E}\kern-.125emX}}
\usepackage{balance}
\usepackage{paralist}
\usepackage{booktabs}
\usepackage{caption}
\usepackage{tabularx}
\usepackage{amsmath} 
\usepackage{amssymb} 
\usepackage{graphicx}
\usepackage{hyperref}
\usepackage{todonotes}
\usepackage[inkscapelatex=false]{svg}
\usepackage{subcaption}   
\usepackage{varwidth}
\usepackage{tabularray}

\usepackage[ruled,vlined]{algorithm2e}
\usepackage{float}
\usepackage{multirow}
\usepackage{adjustbox}
\usepackage{tikz}
\usepackage{cite}
\usepackage{makecell}

\newcommand{\argmin}{\mathop{\mathrm{arg\,min}}\limits}

\newcommand{\giu}[1]{\textcolor{black}{#1}}

\newcommand{\den}{\mathcal{M}_{den}}
\newcommand{\att}{\mathcal{A} (\nabla, f_\theta)}
\newcommand{\tk}{Top-\textit{k}}

\newcommand{\tool}{{\textsc{{GUIDE}}}}

\definecolor{darkgreen}{RGB}{0,100,0}

\newcommand{\ballnumber}[1]{%
  \tikz[baseline={(myanchor.base)}]{%
    \node[circle, draw, fill=none, inner sep=0pt, minimum width=1em, minimum height=1em, text centered] (myanchor) {#1};
  }%
  \ignorespaces
}

\newcommand{\ballnumberspace}[1]{%
    \tikz[baseline={(myanchor.base)}]{
        \node[circle, draw, fill=none, inner sep=0pt, minimum width=1em, minimum height=1em, text centered] (myanchor) {\bfseries\scriptsize#1};
    }%
}

\usepackage[section,numberedsection=autolabel,nogroupskip,nonumberlist]{glossaries}
\makeglossaries
\glsdisablehyper

\newacronym{fl}{FL}{Federated Learning}
\newacronym{gias}{GIAs}{Gradient Inversion Attacks}
\newacronym{bn}{BN}{Batch Normalization}
\newacronym{fedsgd}{FedSGD}{Federated Stochastic Gradient Descent}
\newacronym{fedavg}{FedAVG}{Federated Averaging}
\newacronym{gan}{GAN}{Generative Adversarial Network}
\newacronym{fc}{FC}{Fully Connected}
\newacronym{cnn}{CNN}{Convolutional Neural Network}
\newacronym{dp}{DP}{Differential Privacy}
\newacronym{sa}{SA}{Secure Aggregation}
\newacronym{cdp}{CDP}{Centralized Differential Privacy}
\newacronym{ldp}{LDP}{Local Differential Privacy}
\newacronym{cml}{CML}{Centralized Machine Learning}
\newacronym{ood}{OOD}{Out-Of-Distribution}
\newacronym{ssim}{SSIM}{Structural Similarity}
\newacronym{psnr}{PSNR}{Peak Signal-to-Noise Ratio}
\newacronym{lpips}{LPIPS}{Learned Perceptual Image Patch Similarity}
\newacronym{mse}{MSE}{Mean Square Error}
\newacronym{he}{HE}{Homomorphic Encryption}
\newacronym{fhe}{FHE}{Fully Homomorphic Encryption}
\newacronym{fu}{FU}{Federated Unlearning}
\newacronym{ml}{ML}{Machine Learning}
\newacronym{niid}{non-IID}{Nonindependent and Identically Distributed}
\newacronym{qsgd}{QSGD}{Quantized Stochastic Gradient Descent}

\begin{document}

\title{\tool{}: Enhancing Gradient Inversion Attacks in Federated Learning with Denoising Models}

\author{Vincenzo~Carletti, Pasquale~Foggia, Carlo~Mazzocca, Giuseppe~Parrella, and Mario~Vento%
\thanks{V. Carletti, P. Foggia, C. Mazzocca, G. Parrella, and M. Vento are with the Department of Computer Information and Electrical Engineering and Applied Mathematics, University of Salerno, Via Giovanni Paolo II, 132, 84084 Fisciano, Salerno, Italy (e-mail: \{vcarletti, pfoggia, cmazzocca, gparrella, mvento\}@unisa.it).}
\thanks{* indicates the corresponding author.}
\thanks{Manuscript received \textbf{X}, \textbf{X}; revised \textbf{X}, \textbf{X}.}
}

\markboth{Journal of \LaTeX\ Class Files,~Vol.~18, No.~9, September~2020}%
{How to Use the IEEEtran \LaTeX \ Templates}

\maketitle

\input{sections/01_abstract}
\input{sections/02_introduction}
\input{sections/03_a_background}
\input{sections/03_b_related}
\input{sections/04_threat_model}
\input{sections/05_method}
\input{sections/06_experiments}
\input{sections/07_defensive_measures}

\input{sections/08_conclusions}

\bibliographystyle{IEEEtran}
\bibliography{bibtex}

\end{document}

%% file: sections/01_abstract.tex
\begin{abstract}
Federated Learning (FL) enables collaborative training of Machine Learning (ML) models across multiple clients while preserving their privacy. Rather than sharing raw data, federated clients transmit locally computed updates to train the global model. Although this paradigm should provide stronger privacy guarantees than centralized ML, client updates remain vulnerable to privacy leakage. Adversaries can exploit them to infer sensitive properties about the training data or even to \emph{reconstruct} the original inputs via Gradient Inversion Attacks (GIAs).
\giu{Under the honest-but-curious threat model, GIAs} attempt to reconstruct training data by reversing intermediate updates using optimization-based techniques.
\giu{We observe that these approaches} usually reconstruct \emph{noisy} approximations of the original inputs, whose quality can be enhanced with \emph{specialized} denoising models.
This paper presents Gradient Update Inversion with DEnoising (\tool{}), a novel methodology that leverages diffusion models as denoising tools to improve image reconstruction attacks in FL. \tool{} can be integrated into any GIAs \giu{that exploits surrogate datasets, a widely adopted assumption in GIAs literature.}
We comprehensively evaluate our approach in two \giu{attack scenarios} that use different FL algorithms, models, and datasets. \giu{Our results demonstrate that GUIDE integrates seamlessly with two state-of-the-art GIAs, substantially improving reconstruction quality across multiple metrics. Specifically, GUIDE achieves up to 46\% higher perceptual similarity, as measured by the DreamSim metric.}
\end{abstract}


%% file: sections/02_introduction.tex
\section{Introduction}\label{sec:intro}

\IEEEPARstart{I}{nternet} of Things (IoT) environments are characterized by vast amounts of data that are continuously generated by heterogeneous and resource-constrained devices. While this data is valuable for \gls{ml} applications \cite{9739684}, transmitting it to a central server is often impractical due to privacy, bandwidth, and latency constraints.
Federated Learning (FL)~\cite{mcmahan2017communication} offers a promising alternative to traditional centralized \gls{ml} by enabling multiple devices, such as those in the IoT, to collaboratively train shared models without exchanging the training data~\cite{bellavista2021decentralised}. In an FL setting, a central server distributes a global model to participating clients that locally compute updates (i.e., gradients or weight differences) using their private dataset. The updates are then sent to the server, which aggregates them to obtain an updated version of the global model that is sent back to the clients. This process is repeated until convergence.

Although FL enhances privacy protection compared to centralized approaches as clients' data are kept on-site~\cite{MOTHUKURI2021619}, it is still vulnerable to several security and privacy threats~\cite{survey_inf_fusion, survey_pami,survey_acm}. In particular, privacy attacks, which are the main focus of this work, aim to infer sensitive information about the clients, such as training data or their properties, by exploiting the intermediate updates sent to the server~\cite{survey_pami, survey_acm, oursok, du2024sok}.

In FL scenarios, server-client communications can be protected through end-to-end encryption, making attacks by external adversaries more difficult. However, this does not prevent an \emph{honest-but-curious} server from breaking clients' privacy without being detected, as it follows the regular training protocol while seeking to gain additional information~\cite{survey_pami,oursok}. The server has access to the global model and client updates. Additionally, it is capable of executing computationally intensive reconstruction methods.

Within this threat model, \gls{gias} aim to reconstruct the private client data used to compute the model update~\cite{oursok}. These attacks rely on optimization techniques to reconstruct client's data, with the objective to estimate input values that produce updates closely matching those shared by the victim client~\cite{dlg,invgrad}. Optimization strategies may also employ some pre-trained generative models like \glspl{gan} or diffusion models as strong priors to constrain the optimization process~\cite{gi_gip,ggl,gu_ddpm}. In the latter case, the attacker must have access to a \textit{surrogate dataset} to train these models. These datasets should approximate the target client’s data distribution (e.g., if the learning task consists of recognizing faces, then the attacker should have some public photos of other individuals).
This assumption is widely accepted in the literature~\cite{ggl,gu_ddpm,rog_attack,gifd,hc_dlg,fgla,yao2024urvfl,oursok}.

\begin{figure}[t!]
    \centering
    \begin{adjustbox}{max width=\columnwidth}
        \setlength{\tabcolsep}{5pt} 
        \renewcommand{\arraystretch}{1.2} 
        \begin{tabular}{@{}m{0.05\textwidth}@{}m{0.85\textwidth}@{}}
             \centering (a) & \includegraphics[width=\linewidth]{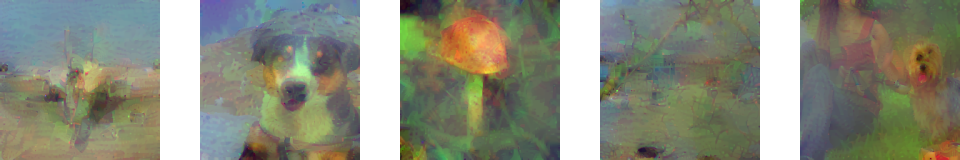} \\
             \centering (b) & \includegraphics[width=\linewidth]{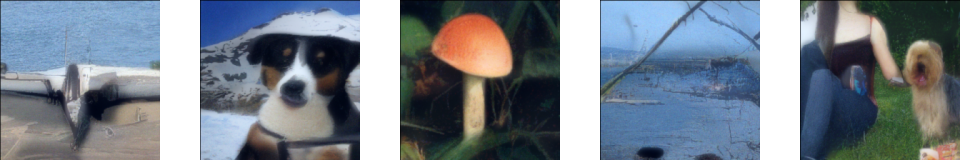} \\
             \centering (c) & \includegraphics[width=\linewidth]{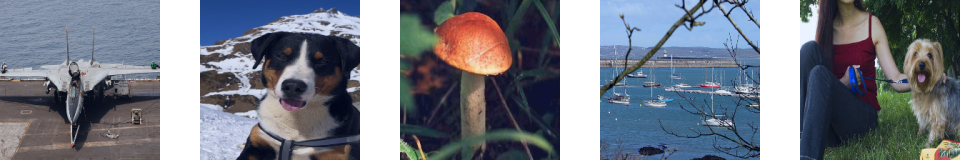} \\
        \end{tabular}
    \end{adjustbox}
    
    \caption{Comparison of reconstruction results on images when the model update is calculated over 256 images at \(224 \times 224\) resolution using the VGG-16 model. \textbf{(a)} Reconstruction with CPA~\cite{cpa_attack}, \textbf{(b)} Reconstruction with CPA~\cite{cpa_attack} enhanced with \tool{} \textit{(ours)}, \textbf{(c)} Original Images.}
    \label{fig:comparison_results}
\end{figure}

\gls{gias} have been demonstrated to be more effective when model updates are computed over small batch sizes or when the data is low-dimensional~\cite{dlg, gradinversion, invgrad, ggl, hc_dlg, gu_ddpm}. However, FL clients usually compute gradients over large batch sizes, and data is high-dimensional (e.g., high-resolution images), making data reconstruction even more challenging. 
As shown in Figure~\ref{fig:comparison_results}a, when larger batches of high-resolution images are used, existing \gls{gias} often generate \emph{noisy} approximations of the client's original private images. 
This limitation can be overcome if the attacker has access to a surrogate dataset to train \textit{tailored} \gls{ml} models aiming to enhance reconstruction quality. To this purpose, recent advancements in generative neural networks, particularly diffusion models~\cite{survey_diff_models}, hold significant promise to make \gls{gias} more effective.

\smallskip
\noindent \textbf{\tool{}.}
Building on these insights, this paper presents \textbf{G}radient \textbf{U}pdate \textbf{I}nversion with \textbf{DE}noising (\tool{}), a novel methodology that leverages diffusion models to enhance the data quality reconstruction of GIAs in FL. By training denoising models for a \textit{specific} reconstruction task, \tool{} maximizes their effectiveness, offering a versatile framework that integrates seamlessly with various attack strategies while preserving their original threat model. \giu{
To the best of our knowledge, this is the first method explicitly designed to enhance the performance of \gls{gias}, while maintaining full compatibility with the underlying attack procedure.
}

We implemented \tool{} and conducted extensive evaluations across various state-of-the-art attacks, using different FL algorithms, models, and datasets, while also considering defensive mechanisms. 
Furthermore, we recognize that commonly used perceptual metrics for assessing \gls{gias} may not be entirely suitable for evaluating privacy leakage in reconstructed images because they do not fully align with human perception~\cite{sem_metrics}. To address this issue, we propose using DreamSim~\cite{dreamsim}, which has recently been established as the state-of-the-art perceptual similarity metric~\cite{lipsim}. Experimental results demonstrate that \tool{} significantly enhances the quality of reconstructed data in \gls{gias}, generating images that are both more realistic and semantically closer to the original ones. Specifically, it achieves up to 46\% higher perceptual similarity according to the DreamSim metric, with even further gains when certain defensive strategies are applied. Figure~\ref{fig:comparison_results}b visually illustrates how \tool{} enhances the effectiveness of the Cocktail Party Attack (CPA)~\cite{cpa_attack}. These improvements, achieved under the same threat model, directly correlate with increased privacy leakage, even when updates are computed over large batch sizes and high-dimensional data.

\smallskip
\noindent \textbf{Contributions.} In the following, we summarize the main contributions of this work:

\begin{itemize}
    \item We explore the use of \textit{specialized} denoising methods to enhance the quality of \gls{gias} in FL;
    \item We propose \tool{}, a novel methodology that enhances the client's input reconstruction in FL by integrating \gls{gias} with denoising mechanisms. To the best of our knowledge, this is the first approach specifically designed to enhance existing GIAs;
    \item We demonstrate that the DreamSim~\cite{dreamsim} metric better assesses the privacy leakage of reconstructed images, offering a more perceptually aligned evaluation of \gls{gias} in FL; 
    \item We implement \tool{} and extensively evaluate it across state-of-the-art attacks under different attack scenarios, demonstrating its effectiveness and adaptability in reconstructing client images. 
\end{itemize}

\smallskip
\noindent \textbf{Organization.} 
The remainder of the paper is organized as follows. Section~\ref{sec:related_work} provides the background and review of related work on \gls{gias}. 
Section~\ref{sec:threat_model} introduces the considered threat model. Section~\ref{sec:methodology} presents \tool{}, whose effectiveness is 
extensively evaluated in Section~\ref{sec:experiments}. Finally, Section~\ref{sec:conclusion} draws our conclusions. Notations and symbols used throughout the paper are clarified in Table~\ref{tab:notation}.

\input{tables/notation}

%% file: tables/notation.tex
\begin{table}[t!]
\centering
\small
\renewcommand{\arraystretch}{1.2} 
\begin{adjustbox}{width=\columnwidth}
\begin{tabular}{lp{6.5cm}} 
\toprule
\textbf{Symbol} & \textbf{Description} \\
\midrule
$\theta$ & Weights of the shared model \\
$f_{\theta}$ & The shared model trained in FL \\
$C$ & Set of federated clients\\
$|C|$ & Cardinality of $C$\\
$c$ & Indexes one of the $C$ clients in the federation \\
$D_c$ & Local private dataset of the client $c$ with $N_c$ samples \\
$B$ & Client local batch size \\
$E$ & Client training epochs \\
$\mathcal{L}$ & Loss function used to train the shared model \\
$\nabla$ & Model update computed by client during FL training \\
$\mathcal{X}$ & Input space (e.g., image space for image classifiers) \\
$\mathcal{Y}$ & Label space \\
$\mathcal{Z}$ & Generative model latent space \\
$\mathcal{L}_{\text{grad}}$ & Loss function to measure gradient similarity \\
$\att$ & Attack procedure\\
$\mathcal{S}_{iters}$ & Set of iteration used to sample noisy data from $\mathcal{A}$ \\
$\mathcal{D}_{iters}$ & Set of iteration in which use denoising model during attack \\
$D'$ & Surrogate dataset owned by attacker\\
$D_{den}$ & Dataset used by the attacker to train denoising model\\
$\den$ & Denoising model\\
\bottomrule
\end{tabular}
\end{adjustbox}
\caption{Formal notations used in the paper.}
\label{tab:notation}
\end{table}

%% file: sections/03_a_background.tex
\section{Background \& Related Work} \label{sec:related_work}
In this section, we provide the background of our work and review state-of-the-art \gls{gias} within the honest-but-curious server threat model, analyzing their methodologies and discussing key limitations. 

\subsection{Federated Learning}
FL is a decentralized learning paradigm that emphasizes data privacy by keeping data stored on client devices. 
FL enables clients to collaboratively train a global model in a privacy-preserving manner as they exchange model updates only, instead of their private training data. 
In a federated setting, each client \( c \) possesses a local dataset \( D_c = \{(x_{c,i}, y_{c,i})\}_{i=1}^{N_c} \), where \( x_{c,i} \in \mathcal{X} \) is an input and \( y_{c,i} \in \mathcal{Y} \) is its corresponding label. 
Each client trains its local model by minimizing the empirical risk over its dataset. 
This local objective for a client \( c \) is expressed as:

\begin{equation}
    f_c(\theta) \triangleq \frac{1}{N_c} \sum_{i=1}^{N_c} \mathcal{L}(x_{c,i}, y_{c,i}; \theta),
\end{equation}
where \( \mathcal{L} \) denotes the loss function used to update the model parameters \( \theta \). 
Given a set of clients \( C \), the main goal in FL is to optimize a global objective function that averages the local objective ones across all participants, as follows:
\begin{equation}
    \min_{\theta} f(\theta) = \frac{1}{|C|} \sum_{c=1}^{|C|} f_c(\theta),
\end{equation}

The primary algorithms used in FL are \gls{fedsgd}~\cite{mcmahan2017communication} and \gls{fedavg}~\cite{mcmahan2017communication}, which mainly differ in how the selected clients \(\mathcal{K} \subseteq C\) compute and transmit updates at each iteration to train the global model. 
In \gls{fedsgd}~\cite{mcmahan2017communication}, each client performs a single local epoch (\(E=1\)) by sampling a batch from its private dataset (\(B \leq N_c\)) and computing the aggregated gradient 
$\nabla_c = \sum_{i=1}^{B} \nabla \mathcal{L}(x_{c,i},y_{c,i};\theta_t)$.
This computed gradient is then sent to the server, which aggregates the gradients from all selected clients to update the global model as follows:
\begin{equation}
\theta_{t+1} \gets \theta_t - \eta \sum_{c=1}^{|\mathcal{K}|} \nabla_c,
\end{equation}
where \( \eta \) is the global learning rate. While \gls{fedsgd} enables frequent global model updates, it incurs significant communication overhead due to the continuous exchange of gradients~\cite{mcmahan2017communication}.  
\gls{fedavg}~\cite{mcmahan2017communication} improves communication efficiency by allowing clients to perform multiple local training steps (\(E>1\)) before sharing updates with the server. In each communication round \( t \), a client \( c \in \mathcal{K} \) initializes its local model as \( \theta_c^{(t,0)} = \theta_t \) and performs \( \tau = E \cdot N_c /B \) gradient descent steps using a learning rate \( \eta \), updating its local model to \( \theta_c^{(t,\tau)} \).  
Once local training is completed, the client computes the model update as:
\begin{equation}
\nabla_c = \theta_c^{(t,0)} - \theta_c^{(t,\tau)},
\end{equation}
and transmits it to the server. The server aggregates the updates by averaging them:
\begin{equation}\label{eq:mod_update}
\theta_{t+1} \gets \frac{1}{|\mathcal{K}|} \sum_{c \in \mathcal{K}} \nabla_c,
\end{equation}
thereby generating the updated global model \( \theta_{t+1} \), which is then distributed to all clients for the next training round. For clarity, Equation~\ref{eq:mod_update} is valid if the number of local samples is the same for each client; if this assumption is not satisfied, the average is replaced by a weighted average, using the number of local samples as the weight.

%% file: sections/03_b_related.tex
\subsection{Gradient Inversion Attacks} 

In the standard setting of FL with an honest-but-curious server, \gls{gias} aim to recover a client’s private training data \(X = \{x_1, \dots, x_n\}\) and labels \(Y = \{y_1, \dots, y_n\}\) starting from the model update $\nabla$ shared by the client during the training procedure. This class of attacks typically formulates input reconstruction as the following optimization problem:

\begin{equation} \label{eq:opt_based_attacks}
\begin{split}
    X^*, Y^* = \arg\min_{X', Y'} \, \mathcal{L}_{\text{grad}} \left(\nabla, \nabla'\right) + \mathcal{R}_{\text{aux}}(X'),
\end{split}
\end{equation}

where $\mathcal{L}_{\text{grad}}$ denotes a loss function 
that quantifies the discrepancy between the shared update $\nabla$ 
and the dummy update $\nabla'$ computed with respect to the reconstructed 
input $X'$ and labels $Y'$. $\mathcal{R}_{\text{aux}}$ is a regularization term that enforces specific constraints on the reconstructed input $X'$. 
The optimization variables \( X' \) and \( Y' \) can be initialized either randomly or leveraging prior knowledge relevant to the problem~\cite{oursok}. 

One of the first attacks is Deep Leakage from Gradients (DLG)~\cite{dlg}, which targets small batches and simple architectures under the \gls{fedsgd} protocol. Subsequent works have demonstrated the effectiveness of similar methods in the presence of more complex FL schemes, such as \gls{fedavg}~\cite{invgrad2, rog_attack}. 
Progress in this domain has generally involved refining the optimization pipeline with additional priors and constraints. For instance, InvGrad~\cite{invgrad} uses a combination of cosine distance and total variation regularization, thereby enabling the inversion of deeper networks and larger batches. GradInversion~\cite{gradinversion} further exploits \gls{bn} statistics to guide reconstructions, aligning the activations produced by candidate inputs with the global \gls{bn} statistics stored in the model. Meanwhile, some works isolate label reconstruction as a separate challenge, which can be addressed analytically (e.g., assuming unique labels per batch). In another direction, CPA~\cite{cpa_attack} casts gradient inversion in \gls{fc} layers as a blind source separation problem, first recovering private embeddings by unmixing the \gls{fc}-layer gradients via techniques akin to independent component analysis, and then reconstructing images through an auxiliary optimization. 

Some other attacks~\cite{gi_gip,ggl,gifd,cgir} leverage pre-trained generative models to reduce the complexity of searching directly in the high-dimensional image space. Concretely, these approaches restrict the optimization process to a latent code \(z' \in \mathcal{Z}\), under the assumption of having a generator \(G \colon \mathcal{Z} \to \mathcal{X}\) that maps latent vectors to candidate images. The server-side updates \(\nabla'\) are then computed by substituting \(G(z')\) into the client’s model in place of the actual private inputs. Formally, one solves:
\begin{equation}\label{eq:latent_space_rec_attacks}
    \begin{aligned}
        z^* \;=\; \argmin_{z' \in \mathcal{Z},\; Y' \in \mathcal{Y}} &\ 
        \mathcal{L}_{\text{grad}} \bigl(\nabla,\;\nabla'\bigr) 
        \;+\; \mathcal{R}_{\text{aux}} \bigl(G;\,z'\bigr),
    \end{aligned}
\end{equation}
where \(\nabla\) denotes the client update and \(\nabla'\) are computed on the generated data \(G(z')\) and labels $Y'$. A fundamental assumption of these approaches is that the attacker has access to a \emph{surrogate dataset} sampled from a distribution similar to that of the victim's data, allowing for effective pre-training of \(G\). Although this latent-space formulation can expedite and stabilize the reconstruction process, it often necessitates additional fine-tuning steps for each image of interest, thereby increasing the overall computational overhead of the attack.

Recently,~\cite{gu_ddpm} have proposed a novel GIA based on the latent-space optimization principle (Eqn.~\ref{eq:latent_space_rec_attacks}), employing a diffusion model as the pre-trained generative model instead of a GAN.
It is worth noting that the diffusion model is used as a generative prior, guiding the iterative synthesis process from pure noise by incorporating the shared gradient as a directive condition at each sampling step. 
This significantly differs from our methodology that leverages the diffusion model as a denoising tool specialized for the underlying GIA. Additionally,~\cite{gu_ddpm} was validated on model updates computed from a single image (a batch size of one), which is an unrealistic setting for practical FL deployments.

\subsection{\giu{Augmented Optimization Attacks}} 
Other strategies seek to improve \gls{gias} by incorporating image restoration modules, particularly when defensive measures based on gradient obfuscation are in place~\cite{oursok}. 
Highly-Compressed Gradient Leakage Attack (HC-GLA)~\cite{hc_dlg} combines standard gradient-based reconstruction steps with a specialized \gls{cnn} for noise suppression. 
This method differs from state-of-the-art image restoration models, as it is a \gls{cnn} trained on a limited dataset of only a few thousand images, which restricts its overall potential and effectiveness in comparison to more advanced approaches.
Moreover, it is primarily evaluated on single-image batches and may struggle to scale or generalize to diverse data distributions or larger batch sizes.

Reconstructing from Obfuscated Gradient (ROG)~\cite{rog_attack} attack introduces a two-step GIA designed to enhance both convergence and visual fidelity. First, the attack operates in a reduced-dimensional space by encoding the input data, simplifying the optimization objective, and accelerating convergence. Once a suitable representation is found, a \gls{gan} is used to restore the optimized embedding into the original high-dimensional image space. Unlike \tool{}, this \gls{gan} is not specifically trained on the intermediate noisy reconstructions produced by the initial attack step. This limits its adaptability and results in a significant performance drop when targeting deeper neural architectures.

\giu{
However, these approaches incorporate image restoration models directly into the attack logic, making them inherently limited and not generalizable to other \gls{gias}. In contrast, \tool{} explicitly integrates denoising models into the optimization process of the underlying GIA, thereby enhancing reconstruction quality in a flexible and attack-agnostic manner. To the best of our knowledge, this is the first method specifically designed to be compatible with, and to improve the performance of, a broad range of \gls{gias}.
}

%% file: sections/04_threat_model.tex
\begin{figure*}[t!]
    \centering
    \includegraphics[width=1.0\textwidth]{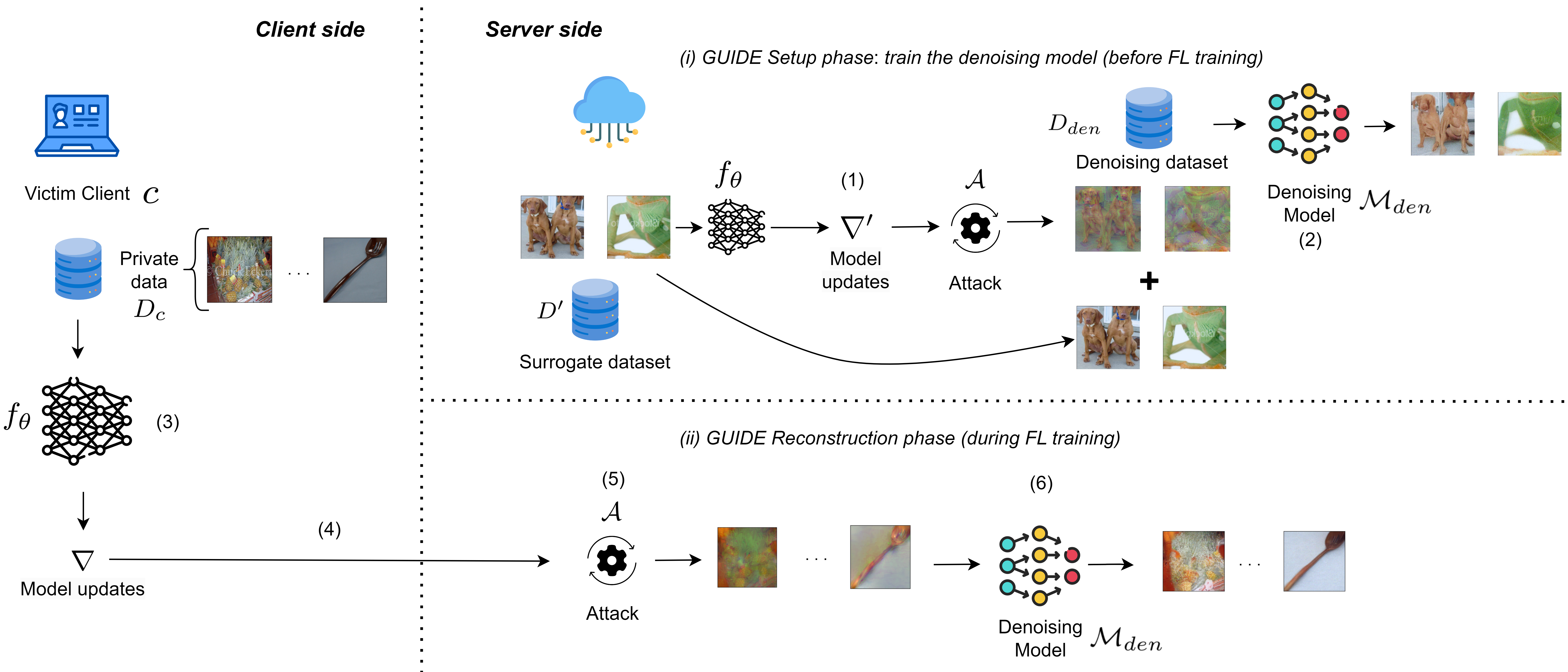} 
    \caption{\tool{} overview. During the setup phase \textit{(i)}, the server constructs a denoising dataset $D_{den}$ by using model updates $\nabla'$ generated from the shared model $f_\theta$, a surrogate dataset $D'$, and the attack procedure $\mathcal{A}$ to produce \textit{noisy} images \textit{(1)}. The attacker then trains a denoising model \textit{(2)} on this newly created dataset $D_{den}$. During the reconstruction phase \textit{(ii)}, the client performs local training steps on private data \textit{(3)}, resulting in a model update $\nabla$ that is sent to the server \textit{(4)}. The attacker then applies the attack procedure $\mathcal{A}$ to generate noisy reconstructions of the client's data \textit{(5)} and subsequently improves the quality of these reconstructions by using the diffusion-based denoising model trained in the earlier phase \textit{(6)}.}
    \label{fig:guide_overview}
\end{figure*}

\section{Threat Model} \label{sec:threat_model}
\giu{In this section, we provide a comprehensive definition of the threat model adopted in our work, specifying the nature of the adversary, along with its objectives, knowledge, and capabilities. In line with the literature on \gls{gias}~\cite{oursok,du2024sok}, we consider the central server as the adversary.}

\smallskip
\noindent \textbf{Adversary Goal.} The main goal of the central server is to break clients' privacy by gaining information about their private data used for the training. 

\smallskip
\noindent \textbf{Adversary Knowledge.} The information available to an honest-but-curious server that can be exploited within an attack strategy is inherently related to its role during the training process in FL, or to other information that can be easily gathered. Specifically, the server has access to:
\begin{compactitem}
    \item The shared model $f_\theta$ and its initialization;
    \item The updates $\nabla$ transmitted by clients during the training process;
    \item The ML task undertaken by clients (e.g., face recognition, object classification);
    \item Knowledge about the properties of the data utilized by clients, such as data type, data structure, and image resolution;
    \item Basic parameters of the training process used by clients, including learning rate, batch size, and number of local iterations in \gls{fedavg};
    \item An auxiliary surrogate dataset $D' = \{ (x'_i, y'_i)
        \}_{i=1}^{n'}$ that resembles the data distribution of the victim
        client. For clarity, we emphasize that this dataset does not need
        to have an actual
        overlap with the victim client's data, but must be representative
        of the client's data distribution. For several learning tasks, suitable
        public datasets are available, enabling the server to obtain such an
        auxiliary data source.
\end{compactitem}

In some prior works, the use of a surrogate dataset is not explicitly stated; however, this assumption is implicitly made through the application of pre-trained generative models. \giu{Specifically, this threat model closely aligns with one of the threat models defined in~\cite{oursok}, followed by several works in literature~\cite{ggl,gu_ddpm,rog_attack,gifd,hc_dlg,fgla,yao2024urvfl}.}

\smallskip
\noindent \textbf{Adversary Capability.} \giu{The central server in FL could be either \textit{active malicious} or \textit{honest-but-curious}~\cite{oursok}. An active malicious server intentionally deviates from the protocol to compromise client privacy by manipulating the global model to facilitate data reconstruction. In contrast, we adopt the honest-but-curious server, a more realistic threat model in FL deployments. This reflects scenarios where the server faithfully follows the FL protocol, but passively attempts to infer private information from clients' updates~\cite{oursok}. It is worth noting that, under this threat model, clients are unable to detect the server's malicious behavior.
}

%% file: sections/05_method.tex
\section{\tool{}}\label{sec:methodology} 

We propose \tool{}, a novel methodology designed to enhance the reconstruction quality of \gls{gias}. In this section, we first outline the intuition behind our work. Then, we provide a detailed description of our methodology, which consists of two key phases: \giu{\textit{(i)}} the Setup phase, where the attacker builds a tailored denoising model, and \giu{\textit{(ii)}} the Reconstruction phase, which allows the attacker to recover victim client's input data.

\subsection{Intuition behind \tool{}}\label{sec:rationale} 

Several works have experimentally shown the effectiveness of \gls{gias} to reconstruct an approximated version of the client's input, especially when additional priors constrain the optimization procedure~\cite{oursok}. However, there is no theoretical guarantee that these reconstructions correspond to the data used by the client during the local training. Indeed, \gls{gias} based on gradient-matching optimization procedures (e.g., those based on Equations~\ref{eq:opt_based_attacks} and~\ref{eq:latent_space_rec_attacks}) are inherently ill-posed mathematical problems since they potentially admit several solutions consistent with the observed client's gradient~\cite{zhu2021rgap}. 

The reconstructed inputs obtained through GIAs often exhibit noisy artifacts,
which can be attributed to the fact that the optimization may place the
reconstructed points in lower-density regions of the data manifold, where the
original client data is less likely to reside. Thus, the recovered points may
satisfy the observed gradients but still deviate from the true data
distribution. Given these considerations, we leverage generative image
enhancement models trained with a surrogate dataset to move approximated input
reconstructions into regions of the data space that are more likely to contain
the original client inputs~\cite{vincent2011}. This \emph{``guides''} the optimization
process toward high-density areas of the learned distribution (hence 
the name of the method), reducing ambiguity in the reconstruction and improving the overall
fidelity of the
recovered samples. 

\input{tables/alg1}

\subsection{GUIDE Setup}\label{sec:guide_setup} 

In the setup phase, the attacker uses a surrogate dataset to train a denoising model, which will later be exploited in the reconstruction phase. This process is illustrated in phase \textit{(i)} of Figure~\ref{fig:guide_overview}. Formally, let $\att$ represent an attack procedure that, given the shared update $\nabla$ from the victim client,  knowledge of the shared model $f_\theta$, and other basics information about local training procedure of clients (e.g., those described in Section~\ref{sec:threat_model}) aims to reconstruct an approximation of the input data $\hat{X} = \{ \hat{x}_1, \ldots, \hat{x}_N \}$. 
The attack procedure $\att$ operates iteratively, refining the estimates of the user's inputs over $T$ iterations.
According to the threat model described in Section~\ref{sec:threat_model}, an attacker can exploit a surrogate dataset $ D' = \{ (x'_i,y'_i) \}_{i=1}^{n'} $ to create a set of \textit{noisy} images through $\mathcal{A}$. In Algorithm~\ref{alg:guide_setup}, the attacker leverages the shared model and the surrogate dataset $D'$ to compute a model update $\nabla'$ over a batch of data $\{ ({x'}_i, y'_i)\}_{i=1}^{N}$ drawn from $D'$ (line~\ref{alg:line:update} of Algorithm~\ref{alg:guide_setup}). The model update, which depends on the learning algorithm used, may consist of gradients or model weights and is computed by the \texttt{compute\_model\_update} function. 

Then, the attacker applies the attack procedure $\mathcal{A}(\nabla', f_\theta)$ to obtain an estimated set of user inputs $\{ \hat{x}_i \}_{i=1}^{N}$. 
As the proposed approach is designed to be independent of the specific attack $\mathcal{A}$, it is not possible to generalize when \tool{} can extract noisy images, as this is closely related to the behavior of $\mathcal{A}$. Therefore, we introduce $\mathcal{S}_{iters}$ in Algorithm~\ref{alg:guide_setup} to formalize the set of iterations at which noisy images are sampled.
In the case $\mathcal{S}_{iters} = \{T\}$, images are extracted only at the end of the attack procedure; otherwise, they are sampled during the attack process. 
By sampling images throughout intermediate iterations, rather than solely at the end of the attack, the attacker is enabled to capture and reconstruct images with different noise patterns at various stages of attack $\mathcal{A}$.
This method enables the attacker to build a denoising dataset $D_{den}$, which consists of pairs of noisy images generated by the reconstruction attack and their corresponding original images from $D'$. Using this dataset, the attacker trains a specialized denoising model \( \den: \mathcal{X} \rightarrow \mathcal{X} \), intended to reduce noise in reconstructed images and improve fidelity. 
Details on the specific denoising model are provided in Section~\ref{sec:den_model}. By the end of the setup phase, the attacker tailors a denoising model ready-to-use for the reconstruction. In this work, the input space \( \mathcal{X} \subseteq \mathbb{R}^{C \times W \times H} \) represents images with \( C \) channels, width \( W \), and height \( H \). Although this study focuses on image data, according to most of the existing attacks, \tool{} is a general methodology potentially extensible to any \gls{gias} in FL.

\subsection{\tool{} Reconstruction}\label{sec:guide_rec} 
This stage involves the integration of the denoising model to enhance the reconstruction achieved by the attack procedure $\mathcal{A}$, as described in Algorithm~\ref{alg:guide_rec}.
The attacker’s first step is to execute the setup phase, thereby obtaining a denoising model $\den$, which will be used to enhance reconstruction quality. Subsequently, the attack can be launched, choosing between the two configurations determined by the set $\mathcal{D}_{iters}$, which contains all iterations at which the denoising model must be applied.
In our work, the denoising model $\den$ is applied \textit{after} the base reconstruction procedure $\att$, i.e. when $\mathcal{D}_{iters} = \{T\}$. In this case, $\att$ operates independently to generate an initial reconstruction of the victim client's data, and $\den$ subsequently refines these noisy reconstructions $\hat{X}$ to produce cleaner versions.

\input{tables/alg2}

\input{tables/table_exp_setup}

\subsection{Denoising Model} \label{sec:den_model} 
Image denoising models are essential tools in image processing, aiming to recover high-quality images from their corrupted versions by reducing or eliminating noise. Traditional denoising methods often rely on the statistical properties of the noise and make assumptions about the underlying clean image structures. However, deep learning-based approaches have revolutionized the field by learning complex mappings from noisy to clean images directly from data, achieving state-of-the-art performance in various image restoration tasks~\cite{survey_diff_neuro}. Among these models, those based on diffusion have recently outperformed other leading models in the field for this class of tasks~\cite{survey_diff_models, survey_diff_arx}.

Diffusion models employ a two-phase process: a forward and a reverse process. In the former, noise is gradually added to an image until it is indistinguishable from Gaussian noise. In the reverse process, the model gradually reduces noise by estimating and removing it at each step, progressively reconstructing the original image from a noisy input to achieve high-quality restoration. These models are usually evaluated in different configurations where the perturbations applied to the corrupted images follow specific patterns (e.g., Gaussian noise)~\cite{survey_diff_arx}.

In our work, we leverage denoising models to enhance the quality of images reconstructed by existing attacks. Specifically, we consider the attack outputs $\att$ as noisy versions of the client's training images used to compute the gradient update $\nabla$. 
As mentioned in Section~\ref{sec:threat_model}, the attacker can use a surrogate dataset $D'$ to train the denoising model, which consists of images similar to those possessed by the clients in the FL setting. The noisy images used for training are generated using the attack $\att$ itself, ensuring that the denoising model learns to invert the specific type of degradation introduced by the attack. 

%% file: tables/alg1.tex
\begin{algorithm}[t!]
\caption{\tool{} Setup}
\label{alg:guide_setup}
\scalebox{0.92}{ 
\begin{minipage}{\columnwidth}
\begin{algorithmic}[1]

\REQUIRE The shared model $f_\theta$, the auxiliary surrogate dataset $D'=\{ x'_i\}_{i=1}^{n'}$, an attack procedure $\mathcal{A}$ that runs in $T$ iterations,  a desired cardinality of the denoising dataset $N_{den}$, and a set of iteration $\mathcal{S}_{iters}$ in which data were sampled from $\mathcal{A}$.

\ENSURE A denoising model $\den$
\vspace{0.5em}
\STATE \textcolor{darkgreen}{\textit{// Denoising dataset collection}}
\STATE $D_{den}=\emptyset$

\WHILE{$| D_{den} |  < N_{den} $}
    \STATE \textcolor{darkgreen}{\textit{// Compute model update with surrogate data}}
    \STATE$\nabla'=$\texttt{compute\_model\_update}$(f_\theta, X',D')$ \label{alg:line:update}
    \STATE \textcolor{darkgreen}{\textit{// Get noisy images}}
    \STATE Initialize images $\{\hat{x}_1, \ldots, \hat{x}_N\}$
    \FOR{$t \in 1, \dots, T$}
            \STATE Update $\{\hat{x}_1, \ldots, \hat{x}_N\}$ according to $\mathcal{A}(\nabla', f_\theta)$
            \IF{$t \in \mathcal{S}_{iters}$}
                \STATE $D_{den} \longleftarrow D_{den} \cup \{(x'_i, \hat{x}_1), \ldots,(x'_{N}, \hat{x}_{N}) \}$~\label{alg:line:noisy}
            \ENDIF
    \ENDFOR
    
\ENDWHILE

\STATE \textcolor{darkgreen}{\textit{// Denoising model training phase}}

\STATE $\den = $\texttt{train\_denoising\_model}$(D_{den})$
\RETURN $\den$.
\end{algorithmic}
\end{minipage} }
\end{algorithm}

%% file: tables/alg2.tex
\begin{algorithm}[t!]
\caption{\tool{} Reconstruction}
\label{alg:guide_rec}
\scalebox{0.92}{ 
\begin{minipage}{\columnwidth}
\begin{algorithmic}[1]
 \REQUIRE Model updates from the victim client $\nabla$ computed over $N$ input data, the shared model $f_\theta$, an attack procedure $\att$ that runs in $T$ iterations, a surrogate dataset $D'$, a sampling iteration set $\mathcal{S}_{iters}$ and a denoising iteration set $\mathcal{D}_{iters}$. 

\ENSURE Reconstructed data $\{\hat{x}_1, \ldots, \hat{x}_N\}$ approximating the inputs that produced $\nabla$.
\vspace{0.5em}

\STATE $\den =$\texttt{GUIDE\_setup}$(f_\theta, D', \mathcal{A}, |D'|, \mathcal{S}_{iters})$

\STATE Initialize images $\{\hat{x}_1, \ldots, \hat{x}_N\}$

\FOR{$t \in 1, \dots, T$}
    \STATE Update $\{\hat{x}_1, \ldots, \hat{x}_N\}$ according to $\att$
    
    \IF{$t \in \mathcal{D}_{iters}$}
        \STATE Apply $\den$ on each updated estimate $\hat{x}_i$:\\ $\hat{x}_i = \den(\hat{x}_i)$
    \ENDIF
\ENDFOR

\RETURN $\{\hat{x}_1, \ldots, \hat{x}_N\}$.
\end{algorithmic}
\end{minipage} }
\end{algorithm}

%% file: tables/table_exp_setup.tex

\begin{table*}[t!]
    \centering
    \renewcommand{\arraystretch}{1.3} 
    \begin{adjustbox}{width=\textwidth}
        \begin{tabular}{c l l l l l l l l l}
            \hline
            \textbf{ID} & \textbf{Attack $\mathcal{A}$} & \textbf{Algorithm} & \textbf{Task} & \textbf{Model} & \textbf{Training} & \textbf{Client Data} & \textbf{Surrogate Data} & \makecell[l]{\textbf{Img.} \\ \textbf{Resolution}} & \textbf{Denoising} \\
            \hline
            \ballnumber{1} & \makecell[l]{CPA \cite{cpa_attack}} & FedSGD & Img. Cls. & \makecell[l]{VGG16} & $E{=}1$, $B{\in} \{4, \dots, 256\}$ & \makecell[l]{ImageNet~\cite{imagenet} \\ (Test)} & \makecell[l]{ImageNet~\cite{imagenet} \\ (Val)} & $224 \times 224$ & \makecell[l]{\cite{convir}, \cite{irsde}, \cite{irsde2}} \\
            \ballnumber{2} & \makecell[l]{ROG \cite{rog_attack}} & FedAVG & Face Rec. & \makecell[l]{ResNet-18} & $E{=}5$, $B{=}16$ & \makecell[l]{VGGFace2~\cite{cao2018vggface2} \\ (Subset)} & \makecell[l]{VGGFace2~\cite{cao2018vggface2} \\ (Non-Overlap IDs) } & $128 \times 128$ & \makecell[l]{\cite{irsde}, \cite{irsde2}} \\
            \hline
        \end{tabular}
    \end{adjustbox}
    \caption{Overview of the attack scenarios and experimental setup, including image resolution.}
    \label{tab:attack_scenarios}
\end{table*}

%% file: sections/06_experiments.tex
\section{Experimental Analysis} \label{sec:experiments}
This section offers an extensive evaluation of \tool{}. First, we detail our experimental setup and investigate how the methodology can be evaluated across two attack scenarios using different FL algorithms, models, and datasets. 
Experimental results demonstrate that \tool{} remarkably enhances GIAs even in FL scenarios characterized by large batch sizes, high-dimensional data \giu{(i.e., 256 images of 224$\times$224 pixel)} \giu{or when FedAVG is used as learning algorithm; conditions that typically reduce the efficacy of GIAs~\cite{du2024sok}.} 
Finally, we evaluate \tool{}'s resilience against defense mechanisms that may prevent the reconstruction, such as Differential Privacy (DP)~\cite{survey_dp} and heuristic gradient compression techniques like \gls{qsgd}~\cite{alistarh2017qsgd} and \tk~\cite{aji2017topk}.
All experiments were conducted using an NVIDIA A100-SXM4-80GB GPU within a high-performance infrastructure. 
For transparency and reproducibility, we will release our source code upon paper acceptance.

\begin{figure*}[t!!!]
    \centering
    \begin{adjustbox}{max width=\textwidth}
        \setlength{\tabcolsep}{5pt} 
        \renewcommand{\arraystretch}{1.2} 
        \begin{tabular}{@{}m{0.05\textwidth}@{}m{0.9\textwidth}@{}}
            \centering (a) & \includegraphics[width=\linewidth]{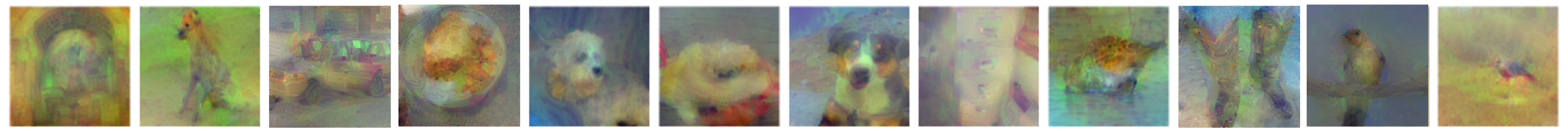} \\
            \centering (b) & \includegraphics[width=\linewidth]{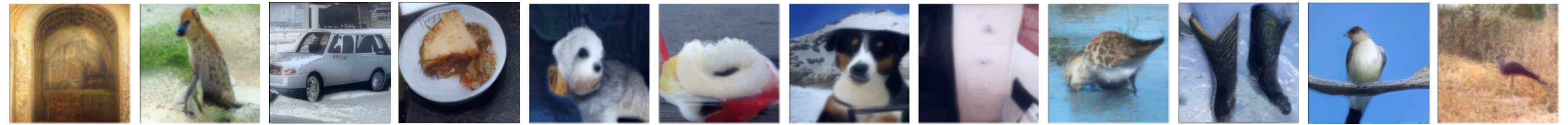} \\
            \centering (c) & \includegraphics[width=\linewidth]{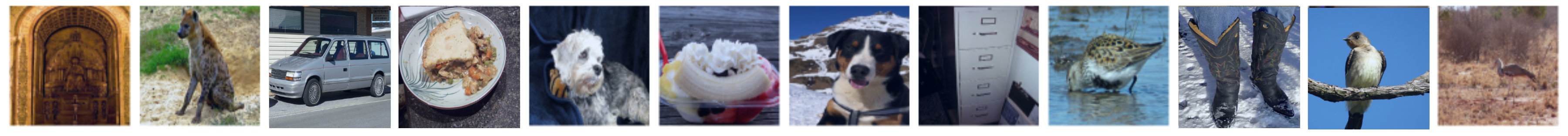} \\
        \end{tabular}
    \end{adjustbox}
    
    \caption{Example of reconstructions obtained by different methods on 12 images in Scenario~\protect\ballnumberspace{1}. The gradient is computed over 256 images with a resolution of $224 \times 224$ with the VGG16 model. All the images are not included in any noisy image seen during the training of the denoising models. \textbf{(a)}~Images reconstructed with CPA~\cite{cpa_attack}. \textbf{(b)}~Reconstruction with CPA~\cite{cpa_attack} enhanced with \tool{} \textit{(ours)}. \textbf{(c)}~Original images used to compute the gradient.}
    \label{fig:guide_results}
\end{figure*}

\subsection{Experimental Setup} 

In this section, we provide a detailed description of the experimental scenarios, evaluation methodology, and metrics used to assess how \tool{} improves the reconstruction quality of GIAs.

\smallskip
\noindent \textbf{Attack Scenarios.} To demonstrate the robustness and generalization of \tool{} in enhancing reconstruction quality, we consider two state-of-the-art attacks by evaluating its performance across two attack scenarios, whose main features are summarized in Table~\ref{tab:attack_scenarios}. These scenarios involve different attack strategies, shared models, and learning algorithms, allowing us to assess the methodology's effectiveness under varying conditions. 
In both cases, the surrogate dataset used by the attacker is strictly non-overlapping with the private client's dataset to ensure a fair and realistic assessment.

In the first scenario, the adversary exploits CPA~\cite{cpa_attack} for an image classification task using a VGG-16~\cite{vgg16} with FedSGD~\cite{mcmahan2017communication}. In the second one, the attacker uses ROG~\cite{rog_attack} for a face recognition task employing ResNet-18~\cite{resnet} with FedAVG~\cite{mcmahan2017communication}. Both scenarios incorporate tailored denoising techniques to improve reconstruction quality. A detailed description of each scenario, along with the denoising strategies employed and the result obtained, is provided in Sections~\ref{sec:attack_scen1} and~\ref{sec:attack_scen2}.

\smallskip
\noindent \textbf{Evaluation Metrics.}  
In the evaluation of \gls{gias} within the computer vision domain, image similarity is typically assessed using metrics such as Peak Signal-to-Noise Ratio (PSNR), Structural Similarity Index Measure (SSIM), and Learned Perceptual Image Patch Similarity (LPIPS)~\cite{lpips}.
PSNR measures the pixel-level fidelity between images, SSIM assesses structural similarities by considering luminance, contrast, and texture, while LPIPS leverages deep neural networks to approximate human perceptual judgments. However, these metrics primarily focus on pixel-wise differences and may not accurately reflect human perception when used to evaluate reconstructed images, particularly in terms of privacy leakage and  recognizability~\cite{rog_attack,sem_metrics}.  

To address these limitations, we also adopt the recently proposed \textit{DreamSim} metric~\cite{dreamsim}, which is designed to better align with human perception. Unlike the above-mentioned, DreamSim evaluates mid-level semantic variations, such as pose, perspective, and object shape, by leveraging high-level features from pre-trained vision models fine-tuned with human-aligned perceptual data. By computing perceptual similarity through cosine distance on feature embeddings, DreamSim offers a more comprehensive assessment of image recognizability, extending beyond pixel-level similarities. To the best of our knowledge, we are the first to use DreamSim for evaluating \gls{gias} in FL, demonstrating its potential to offer a more accurate and human-aligned reconstruction quality assessment.

\subsection[Scenario 1]{Attack Scenario \ballnumber{1} : Image Classification}\label{sec:attack_scen1}

\begin{figure*}[t!]
    \centering
    \includegraphics[width=1.0\textwidth]{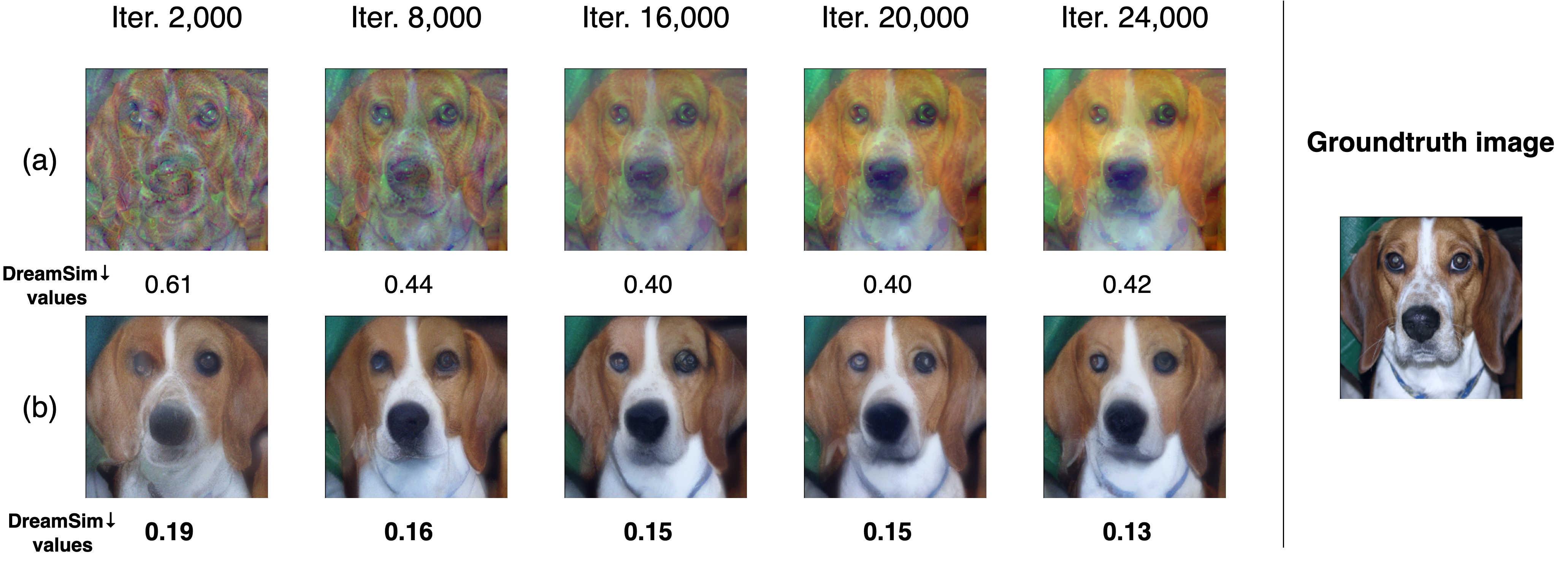}
    \caption{\textbf{(a)} Examples of images generated by CPA~\cite{cpa_attack} at various stages of the attack ($\mathcal{S}_{iters} = \{2000, 8000, 16000, 20000, 24000\}$) when gradients are computed over a batch of 256 images. \textbf{(b)} Reconstructed image obtained by starting from the image of (a) and applying denoising with a blended model. The denoising model demonstrates an ability to enhance reconstruction quality even in the early phases of the attack, as evidenced by lower DreamSim values.}
    \label{fig:early}
\end{figure*}

In this attack scenario, we consider CPA~\cite{cpa_attack} as the base attack $\mathcal{A}$, utilizing a pre-trained VGG-16~\cite{vgg16} as the shared model. The model is trained with FedSGD for one local epoch, with batch sizes ranging from 4 and 256. The victim client's dataset consists of $224\times224$ images from the ImageNet~\cite{imagenet} test set, while the surrogate dataset is a subset of the validation set. To improve reconstruction quality, we employ denoising techniques based on both diffusion models~\cite{irsde,irsde2} and convolutional denoising models~\cite{convir}.

\smallskip \noindent \textbf{\tool{} Setup Phase.}
In this phase, the attacker samples images from the surrogate dataset, computes model updates $\nabla'$ using batch sizes ranging from 4 to 256, obtaining noisy images by applying CPA~\cite{cpa_attack} as the base attack $\mathcal{A}$ to reconstruct the original data. In our experiments, the resulting noisy dataset $D_{den}$ consists of 106,008 pairs of images, each containing a noisy image and its corresponding original version. These images are collected at multiple iterations of 2000 of the CPA to train a denoising model capable of reconstructing images even during the intermediate iterations of the attack.
A diffusion-based model and a convolutional model are employed to denoise the reconstructed images. After collecting the data, we performed an 80\%/20\% split for training and testing and trained both models using the original hyperparameters from~\cite{irsde, convir}. The diffusion model, based on a Conditional NAFNet architecture, was trained for 700,000 iterations with a batch size of 16, using the Lion optimizer~\cite{lion} and optimizing the $L_1$ loss function. The convolutional model, following the \textit{base} architecture from~\cite{convir}, was trained for 300 epochs with a batch size of 64 using the Adam~\cite{adam} optimizer. No data augmentation techniques were applied in either case. Based on the performance evaluation of both models, we selected a combined denoising approach as $\den$, which merges their outputs using a weighted fusion scheme (0.65 for the diffusion model and 0.35 for the convolutional model), as it yielded superior reconstruction quality.

\smallskip \noindent \textbf{\tool{} Reconstruction Phase.}
In this section, we assess the capability of \tool{} to enhance image reconstruction through CPA from various perspectives. \giu{We evaluate GUIDE with $\mathcal{D}_{iters}=\{25000\}$, when denoising is applied after the base CPA attack.}

The performances of \tool{} are evaluated in comparison to CPA without using the denoising models. For all the attacks, the updates are computed over varying numbers of images, ranging from 4 to 256, with all the images sourced from the ImageNet test set. This ensures that the samples to be reconstructed have not been previously encountered in their noisy forms during the training of the denoising model, thereby enabling a realistic evaluation of the models' generalization capabilities.

\input{tables/results_scen1}

\giu{
In Table~\ref{tab:scenario1_results}, we report the results obtained by varying the batch size from 4 to 256 images. Across all batch sizes, \tool{} consistently achieves superior performance in terms of PSNR, with a mean improvement of 18\%, and comparable results for SSIM, where \tool{} registers a mean improvement of 3\%. 
Although the LPIPS metric favors CPA, a visual comparison (see Figure~\ref{fig:guide_results} and~\ref{fig:early}) reveals that images generated by CPA are not semantically closer to the originals. This suggests that LPIPS may not be an appropriate perceptual metric for evaluating privacy leakage in this context, as also discussed in~\cite{sem_metrics}. 
Conversely, DreamSim highlights a substantial mean improvement of 27\% for \tool{} over CPA. This result, further supported by visual inspection, indicates that DreamSim offers a more reliable assessment of privacy leakage in \gls{gias}, in line with its proven effectiveness in other scenarios focused on perceptual similarity~\cite{dreamsim,lipsim}.
}

\smallskip
\noindent \textbf{\tool{} with Early Stopping.} We highlighted that constructing a denoising model trained on images obtained at various stages of a reconstruction attack enables good reconstructions, even when the model is applied to intermediate iterations of the attack. In Figure~\ref{fig:guide_results}, we present the results when \tool{} is applied after the CPA attack, which involves 25,000 iterations in the feature inversion phase. However, given the denoising model’s performance, our approach also holds the promise to achieve better performance than a full CPA attack with fewer iterations. Reducing the number of iterations needed to recover images could result in a more efficient yet potentially dangerous attack as this would significantly lower the computational demands. Moreover, this reduction would allow the attacker to focus more resources on the denoising model’s training phase. Once this model has been trained, it becomes a highly adaptable tool, suitable for deployment across various attack scenarios while minimizing the need for computationally demanding iterative processes.

\begin{table}[t!]
    \centering
    \begin{adjustbox}{width=\columnwidth}
        \begin{tabular}{lccccc}
            \toprule
            \textbf{Attack} & \textbf{Iterations} & \textbf{PSNR}~$\uparrow$ & \textbf{SSIM}~$\uparrow$ & \textbf{LPIPS}~$\downarrow$ & \textbf{DreamSim}~$\downarrow$ \\
            \midrule
            CPA & 25000 & $12.35_{\pm 2.07}$ & \textbf{0.35}$_{\pm 0.14}$ & \textbf{0.48}$_{\pm 0.10}$ & $0.60_{\pm 0.14}$ \\
            \midrule
            GUIDE & & & & & \\
            \quad & 2000 & $13.06_{\pm 2.20}$ & $0.29_{\pm 0.12}$ & $0.66_{\pm 0.08}$ & $0.63_{\pm 0.15}$ \\
            \quad & 4000 & $13.58_{\pm 2.41}$ & $0.31_{\pm 0.13}$ & $0.62_{\pm 0.09}$ & $0.55_{\pm 0.18}$ \\
            \quad & 5000 & $13.91_{\pm 2.35}$ & $0.32_{\pm 0.14}$ & $0.60_{\pm 0.09}$ & $0.52_{\pm 0.17}$ \\
            \quad & 6000 & $13.99_{\pm 2.32}$ & $0.33_{\pm 0.14}$ & $0.60_{\pm 0.10}$ & $0.51_{\pm 0.18}$ \\
            \quad & 8000 & $\textbf{14.15}_{\pm 2.36}$ & $0.34_{\pm 0.14}$ & $0.58_{\pm 0.10}$ & $\textbf{0.50}_{\pm 0.19}$ \\
            \bottomrule
        \end{tabular}
    \end{adjustbox}
    \caption{Performance metrics for CPA~\cite{cpa_attack} and GUIDE across different iterations. Each \tool{} configuration is evaluated over $I$ iterations of the base CPA, with denoising applied only at the final iteration, denoted by $\mathcal{D}_{iters} = \{I\}$. Gradients are computed across 30 batches of 256 images, and results are averaged.}
     
    \label{tab:attack_guidep_results}
\end{table}

Figure~\ref{fig:early} clearly shows that \tool{} reconstructs images with significantly higher fidelity after 2,000 iterations compared to CPA. In this example, the visual findings are corroborated by a 69\% improvement in the DreamSim metric when denoising is applied to the image obtained at iteration 2000. Additionally, Table~\ref{tab:attack_guidep_results} reports all the metrics when different numbers of iterations are used. With 2000 iterations, \tool{} achieves a higher PSNR than CPA, and this is higher as the number of iterations grows. By 8000 iterations, \tool{} reaches the highest PSNR value observed with an improvement of 15\%. The DreamSim metric highlights the effectiveness of our approach as a comparable quality to CPA is achieved at the 2000-iteration mark, and superior reconstructions emerge from 4000 iterations onward. This suggests that \tool{} can yield more perceptually accurate images, even with a low number of iterations. This outcome indicates that attackers could achieve enhanced reconstruction quality with a reduced computational load. The computational burden can be shifted to the training phase of the denoising model, making it a flexible tool for attackers as it can be potentially deployed in multiple attack scenarios.

\subsection[Scenario 2]{Attack Scenario \ballnumber{2} : Face Recognition}\label{sec:attack_scen2}

In this scenario, we consider ROG~\cite{rog_attack} as the base attack $\mathcal{A}$, using ResNet-18~\cite{resnet} as the shared model, trained for a face recognition task. 
The victim client's data consists of $128\times128$ images from a subset of identities in VGGFace2~\cite{cao2018vggface2}, while the surrogate dataset consists of distinct and non-overlapping identities from the same dataset. The denoising techniques employed follow the methods described in~\cite{irsde,irsde2}.

\smallskip
\noindent \textbf{GUIDE Setup.} 
The setup phase is performed by the attacker by sampling data from the surrogate dataset and computing model updates $\nabla'$ using the FedAVG algorithm with 5 local epochs and a batch size of 16 and the learning rate $\eta=5\times 10^{-3}$, following the same setup of the original work~\cite{rog_attack}.  

As described in Section~\ref{sec:related_work}, ROG~\cite{rog_attack} follows a two-step optimization process, first performing the attack in a reduced-dimensional space to simplify optimization and improve convergence, then using restoration modules to reconstruct images in the original high-dimensional space, enhancing perceptual quality.

In this context, \tool{} enables the training of \textit{specialized} restoration models tailored to the specific distortions introduced during the optimization in the reduced space, rather than relying on general-purpose restoration models as presented in the original paper~\cite{rog_attack}. To achieve this, the attacker collects images after the first step of the attack and constructs a denoising dataset, $D_{den}$. In our experiments, we gathered 106,796 images.  

For the denoising process, we employed a diffusion-based denoising model as $\den$. After collecting the data, we performed an 80\%/20\% split for training and testing and trained the model using the original hyperparameters from~\cite{irsde}. The diffusion model, based on a Conditional NAFNet architecture~\cite{irsde}, was trained for 700,000 iterations with a batch size of 4, using the Lion optimizer~\cite{lion} and optimizing the $L_1$ loss function. In this case, the model's input consists of images from the reduced search space ($32 \times 32$ pixels), while the output is in the original image space ($128 \times 128$ pixels). Consequently, the denoising model is trained for a super-resolution task with a scaling factor of 4.

\begin{figure*}[t!!!]
    \centering
    \begin{adjustbox}{max width=\textwidth}
        \setlength{\tabcolsep}{5pt} 
        \renewcommand{\arraystretch}{1.2} 
        \begin{tabular}{@{}m{0.05\textwidth}@{}m{0.9\textwidth}@{}}
            \centering (a) & \includegraphics[width=\linewidth]{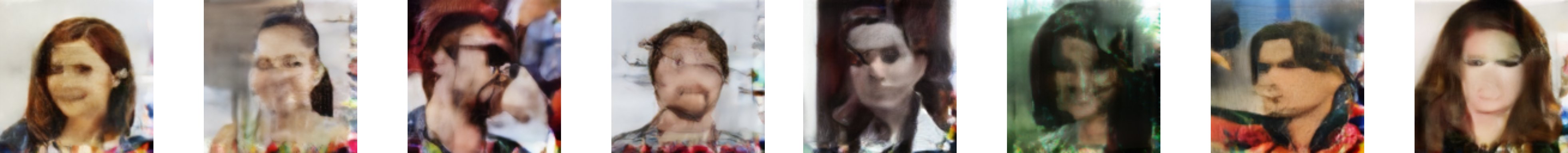} \\   
            \centering (b) & \includegraphics[width=\linewidth]{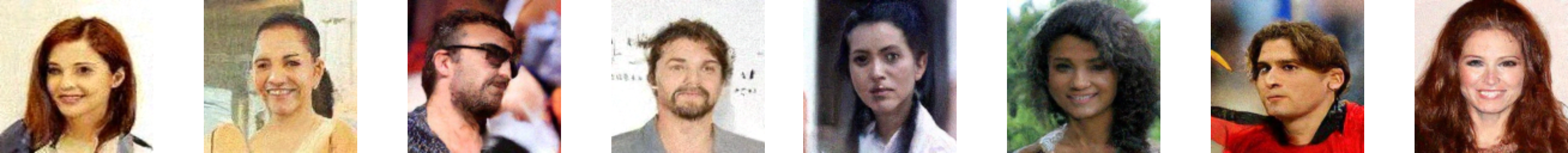} \\   
            \centering (c) & \includegraphics[width=\linewidth]{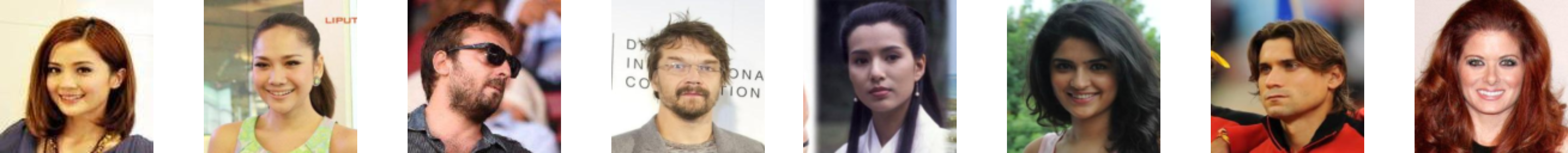} \\   
        \end{tabular}
    \end{adjustbox}
    
    \caption{Example of reconstructions obtained by different methods on 8 images in Scenario~\protect\ballnumberspace{2}. The model update is computed over 16 images with a resolution of $128 \times 128$ with the ResNet-18 model. All the images are not included in any noisy image seen during the training of the denoising models. \textbf{(a)}~Images reconstructed with ROG~\cite{rog_attack}. \textbf{(b)}~Images reconstructed with ROG~\cite{rog_attack} enhanced with GUIDE \textit{(ours)}. \textbf{(c)}~Original images used to compute the gradient.}
    \label{fig:guide_results_scen2}
\end{figure*}

\smallskip
\noindent \textbf{GUIDE Reconstruction.} 
In this section, we assess the capacity of \tool{} to enhance image reconstruction of the ROG~\cite{rog_attack} attack. Specifically, we evaluate the \tool{}, where the denoising model $M_{den}$ is applied at the end of the first optimization step of the attack, with $\mathcal{D}_{iters}=\{100\}$. This choice is motivated by the fact that $M_{den}$ is specifically trained to reconstruct images from the reduced space back to the original input space and, therefore, cannot be directly integrated into the optimization process conducted within the reduced space. Performances are compared with the original ROG configuration presented in~\cite{rog_attack}.

Table~\ref{tab:attack_performance} presents a comparative analysis of reconstruction performances between ROG and \tool{}. The results show that \tool{} consistently outperforms ROG across all evaluated metrics. In particular, \tool{} achieves higher PSNR and SSIM values, indicating improved pixel-wise accuracy and structural similarity with the original images. 
Moreover, \tool{} achieves a lower LPIPS score, indicating superior perceptual quality, and significantly reduces the DreamSim score by 46\%, demonstrating a stronger alignment with human perception.
These quantitative improvements align with the visual results presented in Figure~\ref{fig:guide_results_scen2}, where images reconstructed with \tool{} retain significantly more details than those produced by ROG. In some cases, the enhanced quality of \tool{}’s reconstructions may even enable face recognizability, raising concerns about potential privacy risks for the attacked clients.

\input{tables/results_scen2}
\smallskip
\noindent \textbf{\giu{Robustness to Distribution Shift.}}  
\giu{A natural question that arises is whether the reliance on surrogate data limits the applicability of \tool{}. In this section, we evaluate the effectiveness of our method under distribution shift by employing a surrogate dataset that is not only distinct from the dataset used by the victim client but is also sampled from an entirely different database. Specifically, the denoising model remains trained on the surrogate dataset described in Table~\ref{tab:attack_scenarios} sampled from VGGFace2~\cite{cao2018vggface2}, but the victim client's updates are computed on data from FFHQ~\cite{ffhq}, further increasing the distributional shift between the surrogate and client data.}
This serves as a robustness test to analyze how \tool{} performs when the attacker has access to a visually similar but distinct dataset for training the denoising model while the actual gradients originate from a separate data distribution.
The results in Table~\ref{tab:attack_performance_ffhq} show that \tool{} continues to outperform ROG across all metrics, demonstrating its ability to enhance reconstruction quality even \giu{when the attacker has access to data that comes from a separate distribution with respect to those of the victim client}. In particular, the 32\% reduction in DreamSim suggests that the reconstructed images retain a high degree of perceptual similarity to the original ones, indicating that privacy leakage remains significant despite the increased distributional shift between the client and surrogate datasets.

\input{tables/results_scen2_ffhq}

%% file: tables/results_scen1.tex

\begin{table}[t!]
    \centering
    \begin{adjustbox}{width=\columnwidth}
    \begin{tabular}{clcccc}
        \toprule
        \makecell[c]{\textbf{Batch} \\ \textbf{Size}} & \textbf{Attack} & \textbf{PSNR} $\uparrow$ & \textbf{SSIM} $\uparrow$ 
        & \textbf{LPIPS} $\downarrow$ & \textbf{DreamSim} $\downarrow$ \\
        \midrule
        \multirow{2}{*}{4} 
        & CPA & $12.68_{\pm 2.12}$ & $0.34_{\pm 0.13}$ & $\mathbf{0.47_{\pm 0.10}}$ & $0.58_{\pm 0.13}$ \\
        & GUIDE & $\mathbf{15.10_{\pm 2.20}}$ & $\mathbf{0.35_{\pm 0.14}}$ & $0.53_{\pm 0.08}$ & $\mathbf{0.40_{\pm 0.14}}$ \\
        \midrule
        \multirow{2}{*}{8} 
        & CPA & $12.55_{\pm 2.13}$ & $0.35_{\pm 0.14}$ & $\mathbf{0.49_{\pm 0.12}}$ & $0.60_{\pm 0.15}$ \\
        & GUIDE & $\mathbf{14.84_{\pm 2.62}}$ & $0.35_{\pm 0.15}$ & $0.55_{\pm 0.11}$ & $\mathbf{0.45_{\pm 0.19}}$ \\
        \midrule
        \multirow{2}{*}{16} 
        & CPA & $12.61_{\pm 2.06}$ & $0.33_{\pm 0.14}$ & $\mathbf{0.49_{\pm 0.11}}$ & $0.60_{\pm 0.14}$ \\
        & GUIDE & $\mathbf{14.69_{\pm 2.53}}$ & $\mathbf{0.35_{\pm 0.15}}$ & $0.55_{\pm 0.11}$ & $\mathbf{0.44_{\pm 0.18}}$ \\
        \midrule
        \multirow{2}{*}{32} 
        & CPA & $12.42_{\pm 2.13}$ & $0.32_{\pm 0.13}$ & $\mathbf{0.50_{\pm 0.12}}$ & $0.61_{\pm 0.15}$ \\
        & GUIDE & $\mathbf{14.44_{\pm 2.59}}$ & $\mathbf{0.34_{\pm 0.15}}$ & $0.55_{\pm 0.11}$ & $\mathbf{0.45_{\pm 0.20}}$ \\
        \midrule
        \multirow{2}{*}{64} 
        & CPA & $12.55_{\pm 2.15}$ & $0.34_{\pm 0.15}$ & $\mathbf{0.48_{\pm 0.10}}$ & $0.60_{\pm 0.14}$ \\
        & GUIDE & $\mathbf{14.81_{\pm 2.57}}$ & $\mathbf{0.35_{\pm 0.16}}$ & $0.55_{\pm 0.10}$ & $\mathbf{0.44_{\pm 0.18}}$ \\
        \midrule
        \multirow{2}{*}{128} 
        & CPA & $12.35_{\pm 2.17}$ & $0.33_{\pm 0.14}$ & $\mathbf{0.49_{\pm 0.10}}$ & $0.60_{\pm 0.14}$ \\
        & GUIDE & $\mathbf{14.53_{\pm 2.62}}$ & $\mathbf{0.34_{\pm 0.15}}$ & $0.55_{\pm 0.10}$ & $\mathbf{0.44_{\pm 0.18}}$ \\
        \midrule
        \multirow{2}{*}{256} 
        & CPA & $12.35_{\pm 2.07}$ & $0.35_{\pm 0.14}$ & $\mathbf{0.48_{\pm 0.10}}$ & $0.60_{\pm 0.14}$ \\
        & GUIDE & $\mathbf{14.78_{\pm 2.49}}$ & $\mathbf{0.36_{\pm 0.15}}$ & $0.54_{\pm 0.10}$ & $\mathbf{0.44_{\pm 0.18}}$ \\
        \bottomrule
    \end{tabular}
    \end{adjustbox}
    \caption{Performance comparison of attacks when model update is computed across different batch sizes in Scenario~\protect\ballnumberspace{1}. For each batch size, results are averaged over 30 different batches and reported as $mean_{\pm standard\;deviation}$. The best performance for each metric is highlighted in bold.}
    \label{tab:scenario1_results}
\end{table}

%% file: tables/results_scen2.tex
\begin{table}[t!]
    \centering
    \begin{adjustbox}{width=\columnwidth}
    \begin{tabular}{lcccc}
        \toprule
        \textbf{Attack} & \textbf{PSNR} $\uparrow$ & \textbf{SSIM} $\uparrow$ 
        & \textbf{LPIPS} $\downarrow$ & \textbf{DreamSim} $\downarrow$ \\
        \midrule
        ROG & $17.78_{\pm 1.52}$ & $0.84_{\pm 0.06}$ & $0.51_{\pm 0.05}$ & $0.56_{\pm 0.08}$ \\
        GUIDE & $\textbf{19.50}_{\pm 2.04}$ & $\textbf{0.89}_{\pm 0.06}$ & $\textbf{0.40}_{\pm 0.06}$ & $\textbf{0.30}_{\pm 0.07}$ \\
        \bottomrule
    \end{tabular}
    \end{adjustbox}
    \caption{Performance comparison of attacks in Scenario \protect\ballnumberspace{2} when client data are taken from VGGFace2~\cite{cao2018vggface2}. Updates are computed across 30 batches of 16 images, and results are averaged.}
    \label{tab:attack_performance}
\end{table}

%% file: tables/results_scen2_ffhq.tex
\begin{table}[t!]
    \centering
    \begin{adjustbox}{width=\columnwidth}
    \begin{tabular}{lcccc}
        \toprule
        \textbf{Attack} & \textbf{PSNR} $\uparrow$ & \textbf{SSIM} $\uparrow$ 
        & \textbf{LPIPS} $\downarrow$ & \textbf{DreamSim} $\downarrow$ \\
        \midrule
        ROG & $16.92_{\pm 2.06}$ & $0.79_{\pm 0.15}$ & $0.55_{\pm 0.08}$ & $0.54_{\pm 0.10}$ \\
        GUIDE & $\textbf{18.35}_{\pm 2.93}$ & $\textbf{0.83}_{\pm 0.17}$ & $\textbf{0.42}_{\pm 0.09}$ & $\textbf{0.37}_{\pm 0.12}$ \\
        \bottomrule
    \end{tabular}
    \end{adjustbox}
    \caption{Performance comparison of attacks in Scenario \protect\ballnumberspace{2} when client data are taken from FFHQ~\cite{ffhq}. Updates are computed across 30 batches of 16 images, and results are averaged.}
    \label{tab:attack_performance_ffhq}
\end{table}

%% file: sections/07_defensive_measures.tex
\subsection{Defensive Measures}\label{sec:defensive_measures}

\input{tables/results_scen1_dp}

As \gls{gias} advances, various defensive measures have been developed to protect client privacy. Since the client updates $\nabla$ shared during training are the primary target of \gls{gias}, multiple strategies have been proposed to prevent attackers from accessing and exploiting this information.
The defensive measures can be broadly categorized into cryptographic approaches, formal perturbation methods, and heuristic-based perturbations. Cryptographic techniques, such as \gls{he}~\cite{survey_he} and \gls{sa}~\cite{bonawitz2017practical,sok_sa}, can be integrated into FL training to ensure strong privacy guarantees by preventing direct access to updates. However, these mechanisms introduce significant computational and communication overhead~\cite{sok_sa,he_opt}, limiting their practicality in resource-constrained FL deployments. 
An alternative strategy involves perturbing updates before they are shared. This can be achieved through \gls{dp}~\cite{survey_dp}, which adds carefully calibrated noise to updates based on a rigorous mathematical framework, or with heuristic-based perturbation methods, which modify gradients in an ad-hoc manner to reduce the privacy risk. While these perturbation techniques may counter \gls{gias}, they inherently impact model accuracy and utility due to the noise introduced into the training process.

In this section, we analyze the effectiveness of defensive measures and assess whether \tool{} remains effective in enhancing reconstruction quality when they are applied.  Specifically, we apply privacy-preserving techniques to both attack scenarios considered in this work. To ensure a fair and consistent assessment, we carefully tune these defenses using the parameters established in CPA~\cite{cpa_attack} and ROG~\cite{rog_attack}.  

It is worth noting that as \tool{} integrates advanced, tailored denoising models into the reconstruction process of existing \gls{gias} to improve their reconstruction quality, our methodology is viable as long as the base attack $\mathcal{A}$ generates images that, despite being degraded by noise, retain some similarities with the original data. This allows a trained denoising model to restore image details. However, if the base attack $\mathcal{A}$ fails to produce a recognizable noisy version of the target image, the denoising model becomes ineffective. Consequently, the applicability of \tool{} depends on the extent to which the base attack preserves meaningful features of the original data under different defense mechanisms. Furthermore, an attacker can exploit denoising models more strategically by leveraging knowledge of the specific defense mechanism employed by the client. In such cases, a dedicated denoising model $\den$ can be trained to reconstruct images that have undergone a particular perturbation scheme, thereby adapting to the applied defense and improving the attack's effectiveness.

\input{tables/results_scen2_defenses}

\smallskip \noindent \textbf{Attack Scenario \ballnumber{1} : Image Classification with Differential Privacy.}
This set of experiments demonstrates that when the base attack $\mathcal{A}$ fails to reconstruct a meaningful noisy version of the images, \tool{} does not provide any improvement in reconstruction quality. Table~\ref{tab:dp_results_metrics_comparison} shows that GUIDE fails to enhance reconstruction quality under increasing differentially private noise levels. This is due to the limitations of the CPA attack itself: when CPA produces highly degraded reconstructions, the denoising model in GUIDE lacks sufficient structural information to restore meaningful details. As the noise $\sigma$ increases, both methods exhibit lower PSNR and SSIM, while LPIPS and DreamSim scores worsen, confirming that GUIDE cannot improve reconstruction quality when the base attack fails to retain recognizable image features. The $\sigma$ values used in our evaluation were chosen based on the parameters reported in~\cite{cpa_attack}, ensuring a fair and consistent comparison with prior work.

\smallskip \noindent \textbf{Attack Scenario \ballnumber{2} : Face Recognition with Gradient Perturbation Techniques.}
In this section, we demonstrate that when the base attack $\mathcal{A}$ produces meaningful noisy images, the \tool{} approach can effectively enhance reconstruction quality even in the presence of defensive measures, without altering the underlying threat model.  
By using ROG~\cite{rog_attack} as the base attack, we assess the impact of the following gradient compression techniques, using the same parameter configuration of the original paper~\cite{rog_attack}:  \textit{(i)} \gls{qsgd}~\cite{alistarh2017qsgd} with a 3-bit quantizer; and \textit{(ii)} \tk~\cite{aji2017topk} with a sparsity parameter of 0.95.  
To adapt to these defensive measures, the attacker follows the same procedure described in Section~\ref{sec:attack_scen2} to train two defense-specific denoising models using the surrogate dataset. The training pipeline remains unchanged, except that the dataset $D_{den}$ is now constructed by considering the effects of these defenses, ensuring that the noisy images reflect the perturbations applied on the client side.  
In our experiments, we collect 81,052 images by applying the \gls{qsgd} defense and 76,485 images by applying the \tk ~defense. To obtain specialized denoising models, $\den^{\text{qsgd}}$ and $\den^{\text{Top-\textit{k}}}$, we fine-tune the model previously trained in Section~\ref{sec:attack_scen2}, ensuring it is optimized for reconstructing images under the specific distortions introduced by each defensive measure.

The results in Table~\ref{tab:attack_performance_defenses} demonstrate that \tool{} consistently enhances reconstruction quality across both gradient perturbation defenses.
For the \tk ~defense, \tool{} achieves a 49.15\% reduction in DreamSim, indicating that the reconstructed images are not only sharper but also more semantically aligned with human perception. This suggests an increased risk of privacy leakage despite the applied defense.
For \gls{qsgd}, \tool{} improves reconstruction quality across all evaluated metrics, though to a lesser extent than with \tk. The 29.33\% reduction in DreamSim suggests that even under aggressive quantization, \tool{} enhances perceptual similarity, making reconstructed images more recognizable.
Overall, these findings indicate that while gradient compression techniques reduce reconstruction quality, they do not fully prevent privacy leakage. The ability of \tool{} to adapt by training defense-specific denoising models demonstrates that attackers can partially bypass these defenses, particularly when the base attack retains structural information about the original images.

%% file: tables/results_scen1_dp.tex
\begin{table*}[t!]
    \centering
    \scriptsize
    \begin{adjustbox}{width=\textwidth}
        \begin{tabular}{lcccccccc}
            \toprule
            \multirow{2}{*}{\textbf{$\sigma$ value}} & \multicolumn{2}{c}{\textbf{PSNR} $\uparrow$} & \multicolumn{2}{c}{\textbf{SSIM} $\uparrow$} 
            & \multicolumn{2}{c}{\textbf{LPIPS} $\downarrow$} & \multicolumn{2}{c}{\textbf{DreamSim} $\downarrow$} \\
            \cmidrule(lr){2-3} \cmidrule(lr){4-5} \cmidrule(lr){6-7} \cmidrule(lr){8-9}
            & \textbf{GUIDE} & \textbf{CPA} & \textbf{GUIDE} & \textbf{CPA} 
            & \textbf{GUIDE} & \textbf{CPA} & \textbf{GUIDE} & \textbf{CPA} \\
            \midrule
            $0.0001$ & $\textbf{13.25}_{\pm 2.22}$ & $12.11_{\pm 2.09}$ & $0.28_{\pm 0.13}$ & $\textbf{0.31}_{\pm 0.13}$ & $0.65_{\pm 0.10}$ & $\textbf{0.57}_{\pm 0.10}$ & $0.61_{\pm 0.19}$ & $0.73_{\pm 0.11}$ \\
            $0.001$  & $10.37_{\pm 1.67}$ & $\textbf{10.95}_{\pm 1.91}$ & $0.20_{\pm 0.09}$ & $\textbf{0.20}_{\pm 0.06}$ & $0.79_{\pm 0.06}$ & $\textbf{0.77}_{\pm 0.04}$ & $\textbf{0.84}_{\pm 0.07}$ & $0.87_{\pm 0.06}$ \\
            $0.01$   & $10.38_{\pm 1.70}$ & $\textbf{10.94}_{\pm 1.89}$ & $0.20_{\pm 0.08}$ & $\textbf{0.20}_{\pm 0.06}$ & $0.79_{\pm 0.06}$ & $\textbf{0.77}_{\pm 0.04}$ & $\textbf{0.84}_{\pm 0.06}$ & $0.87_{\pm 0.06}$ \\
            \bottomrule
        \end{tabular}
    \end{adjustbox}
    \caption{Performance comparison of GUIDE and CPA~\cite{cpa_attack} attacks in Scenario \protect\ballnumberspace{1} under differentially private noise perturbation. Results are averaged over 30 batches of 8 images for each metric and $\sigma$ value.}
    \label{tab:dp_results_metrics_comparison}
\end{table*}

%% file: tables/results_scen2_defenses.tex
\begin{table*}[t!]
    \centering
    \scriptsize
    \begin{adjustbox}{width=\textwidth}
        \begin{tabular}{lcccccccc}
            \toprule
            \multirow{2}{*}{\textbf{Defense}} & \multicolumn{2}{c}{\textbf{PSNR} $\uparrow$} & \multicolumn{2}{c}{\textbf{SSIM} $\uparrow$} 
            & \multicolumn{2}{c}{\textbf{LPIPS} $\downarrow$} & \multicolumn{2}{c}{\textbf{DreamSim} $\downarrow$} \\
            \cmidrule(lr){2-3} \cmidrule(lr){4-5} \cmidrule(lr){6-7} \cmidrule(lr){8-9}
            & \textbf{ROG} & \textbf{GUIDE} & \textbf{ROG} & \textbf{GUIDE} 
            & \textbf{ROG} & \textbf{GUIDE} & \textbf{ROG} & \textbf{GUIDE} \\
            \midrule
            \tk~\cite{aji2017topk}  & $16.94_{\pm 1.73}$ & $\textbf{19.16}_{\pm 1.87}$ & $0.81_{\pm 0.07}$ & $\textbf{0.88}_{\pm 0.07}$ & $0.54_{\pm 0.06}$ & $\textbf{0.41}_{\pm 0.06}$ & $0.59_{\pm 0.08}$ & $\textbf{0.30}_{\pm 0.08}$ \\
            QSGD~\cite{alistarh2017qsgd}   & $14.38_{\pm 1.28}$ & $\textbf{15.17}_{\pm 1.54}$ & $0.65_{\pm 0.10}$ & $\textbf{0.73}_{\pm 0.11}$ & $0.68_{\pm 0.05}$ & $\textbf{0.45}_{\pm 0.10}$ & $0.75_{\pm 0.05}$ & $\textbf{0.53}_{\pm 0.07}$ \\
            \bottomrule
        \end{tabular}
    \end{adjustbox}
    \caption{Performance comparison of ROG~\cite{rog_attack} and \tool{} attacks in Scenario \protect\ballnumberspace{2} under different gradient perturbation measures. Results are averaged over 30 batches.}
    \label{tab:attack_performance_defenses}
\end{table*}

%% file: sections/08_conclusions.tex
\section{Conclusion}\label{sec:conclusion}

FL offers a promising way to collaboratively train machine learning models while preserving privacy. However, GIAs can exploit model updates to reconstruct private training data, breaking the client's privacy. This paper proposes \tool{}, a novel methodology tailored to enhance the image reconstruction capabilities of GIAs by leveraging diffusion models as a denoising mechanism. We extensively evaluate \tool{} across state-of-the-art attacks, considering various FL algorithms, models, and datasets.  Experimental results demonstrate that \tool{} can efficiently reconstruct images with high perceptual similarity to the originals, significantly enhancing existing attacks even in challenging settings with large batch sizes and high-dimensional data, and defense mechanisms in place.

%% file: main.bbl
\begin{thebibliography}{10}
\providecommand{\url}[1]{#1}
\csname url@samestyle\endcsname
\providecommand{\newblock}{\relax}
\providecommand{\bibinfo}[2]{#2}
\providecommand{\BIBentrySTDinterwordspacing}{\spaceskip=0pt\relax}
\providecommand{\BIBentryALTinterwordstretchfactor}{4}
\providecommand{\BIBentryALTinterwordspacing}{\spaceskip=\fontdimen2\font plus
\BIBentryALTinterwordstretchfactor\fontdimen3\font minus \fontdimen4\font\relax}
\providecommand{\BIBforeignlanguage}[2]{{%
\expandafter\ifx\csname l@#1\endcsname\relax
\typeout{** WARNING: IEEEtran.bst: No hyphenation pattern has been}%
\typeout{** loaded for the language `#1'. Using the pattern for}%
\typeout{** the default language instead.}%
\else
\language=\csname l@#1\endcsname
\fi
#2}}
\providecommand{\BIBdecl}{\relax}
\BIBdecl

\bibitem{9739684}
J.~Bian, A.~A. Arafat, H.~Xiong, J.~Li, L.~Li, H.~Chen, J.~Wang, D.~Dou, and Z.~Guo, ``Machine learning in real-time internet of things (iot) systems: A survey,'' \emph{IEEE Internet of Things Journal}, vol.~9, no.~11, pp. 8364--8386, 2022.

\bibitem{mcmahan2017communication}
B.~McMahan, E.~Moore, D.~Ramage, S.~Hampson, and B.~A. y~Arcas, ``Communication-efficient learning of deep networks from decentralized data,'' in \emph{Artificial intelligence and statistics}.\hskip 1em plus 0.5em minus 0.4em\relax PMLR, 2017, pp. 1273--1282.

\bibitem{bellavista2021decentralised}
P.~Bellavista, L.~Foschini, and A.~Mora, ``{Decentralised Learning in Federated Deployment Environments: A System-Level Survey},'' \emph{ACM Computing Surveys (CSUR)}, vol.~54, no.~1, pp. 1--38, 2021.

\bibitem{MOTHUKURI2021619}
V.~Mothukuri, R.~M. Parizi, S.~Pouriyeh, Y.~Huang, A.~Dehghantanha, and G.~Srivastava, ``A survey on security and privacy of federated learning,'' \emph{Future Generation Computer Systems}, vol. 115, pp. 619--640, 2021.

\bibitem{survey_inf_fusion}
\BIBentryALTinterwordspacing
N.~Rodríguez-Barroso, D.~Jiménez-López, M.~V. Luzón, F.~Herrera, and E.~Martínez-Cámara, ``Survey on federated learning threats: Concepts, taxonomy on attacks and defences, experimental study and challenges,'' \emph{Information Fusion}, vol.~90, pp. 148--173, 2023. [Online]. Available: \url{https://www.sciencedirect.com/science/article/pii/S1566253522001439}
\BIBentrySTDinterwordspacing

\bibitem{survey_pami}
K.~N. Kumar, C.~K. Mohan, and L.~R. Cenkeramaddi, ``The impact of adversarial attacks on federated learning: A survey,'' \emph{IEEE Transactions on Pattern Analysis and Machine Intelligence}, vol.~46, no.~5, pp. 2672--2691, 2024.

\bibitem{survey_acm}
\BIBentryALTinterwordspacing
X.~Yin, Y.~Zhu, and J.~Hu, ``A comprehensive survey of privacy-preserving federated learning: A taxonomy, review, and future directions,'' \emph{ACM Comput. Surv.}, vol.~54, no.~6, jul 2021. [Online]. Available: \url{https://doi.org/10.1145/3460427}
\BIBentrySTDinterwordspacing

\bibitem{oursok}
\BIBentryALTinterwordspacing
V.~Carletti, P.~Foggia, C.~Mazzocca, G.~Parrella, and M.~Vento, ``Sok: Gradient inversion attacks in federated learning,'' in \emph{34th USENIX Security Symposium (USENIX Security 25)}.\hskip 1em plus 0.5em minus 0.4em\relax USENIX Association, 2025, pp. 6439--6459. [Online]. Available: \url{https://www.usenix.org/conference/usenixsecurity25/presentation/carletti}
\BIBentrySTDinterwordspacing

\bibitem{du2024sok}
\BIBentryALTinterwordspacing
J.~Du, J.~Hu, Z.~Wang, P.~Sun, N.~Z. Gong, K.~Ren, and C.~Chen, ``Sok: On gradient leakage in federated learning,'' 2025. [Online]. Available: \url{https://www.usenix.org/conference/usenixsecurity25/presentation/du}
\BIBentrySTDinterwordspacing

\bibitem{dlg}
L.~Zhu, Z.~Liu, and S.~Han, ``Deep leakage from gradients,'' \emph{Advances in neural information processing systems}, vol.~32, 2019.

\bibitem{invgrad}
J.~Geiping, H.~Bauermeister, H.~Dr\"{o}ge, and M.~Moeller, ``Inverting gradients - how easy is it to break privacy in federated learning?'' in \emph{Proceedings of the 34th International Conference on Neural Information Processing Systems}, ser. NIPS '20.\hskip 1em plus 0.5em minus 0.4em\relax Red Hook, NY, USA: Curran Associates Inc., 2020.

\bibitem{gi_gip}
J.~Jeon, K.~Lee, S.~Oh, J.~Ok \emph{et~al.}, ``Gradient inversion with generative image prior,'' \emph{Advances in neural information processing systems}, vol.~34, pp. 29\,898--29\,908, 2021.

\bibitem{ggl}
Z.~Li, J.~Zhang, L.~Liu, and J.~Liu, ``Auditing privacy defenses in federated learning via generative gradient leakage,'' in \emph{Proceedings of the IEEE/CVF Conference on Computer Vision and Pattern Recognition}, 2022, pp. 10\,132--10\,142.

\bibitem{gu_ddpm}
\BIBentryALTinterwordspacing
H.~Gu, X.~Zhang, J.~Li, H.~Wei, B.~Li, and X.~Huang, ``Federated learning vulnerabilities: Privacy attacks with denoising diffusion probabilistic models,'' in \emph{Proceedings of the ACM Web Conference 2024}, ser. WWW '24.\hskip 1em plus 0.5em minus 0.4em\relax New York, NY, USA: Association for Computing Machinery, 2024, p. 1149–1157. [Online]. Available: \url{https://doi.org/10.1145/3589334.3645514}
\BIBentrySTDinterwordspacing

\bibitem{rog_attack}
\BIBentryALTinterwordspacing
K.~Yue, R.~Jin, C.-W. Wong, D.~Baron, and H.~Dai, ``Gradient obfuscation gives a false sense of security in federated learning,'' in \emph{32nd USENIX Security Symposium (USENIX Security 23)}.\hskip 1em plus 0.5em minus 0.4em\relax Anaheim, CA: USENIX Association, Aug. 2023, pp. 6381--6398. [Online]. Available: \url{https://www.usenix.org/conference/usenixsecurity23/presentation/yue}
\BIBentrySTDinterwordspacing

\bibitem{gifd}
H.~Fang, B.~Chen, X.~Wang, Z.~Wang, and S.-T. Xia, ``Gifd: A generative gradient inversion method with feature domain optimization,'' in \emph{Proceedings of the IEEE/CVF International Conference on Computer Vision}, 2023, pp. 4967--4976.

\bibitem{hc_dlg}
H.~Yang, M.~Ge, K.~Xiang, and J.~Li, ``Using highly compressed gradients in federated learning for data reconstruction attacks,'' \emph{IEEE Transactions on Information Forensics and Security}, vol.~18, pp. 818--830, 2023.

\bibitem{fgla}
D.~Xue, H.~Yang, M.~Ge, J.~Li, G.~Xu, and H.~Li, ``Fast generation-based gradient leakage attacks against highly compressed gradients,'' in \emph{IEEE INFOCOM 2023 - IEEE Conference on Computer Communications}, 2023, pp. 1--10.

\bibitem{yao2024urvfl}
D.~Yao, S.~Li, X.~Gong, S.~Hou, and G.~Pan, ``Urvfl: Undetectable data reconstruction attack on vertical federated learning,'' in \emph{32nd Annual Network and Distributed System Security Symposium, {NDSS} 2025}, 2025.

\bibitem{cpa_attack}
S.~Kariyappa, C.~Guo, K.~Maeng, W.~Xiong, G.~E. Suh, M.~K. Qureshi, and H.-H.~S. Lee, ``Cocktail party attack: Breaking aggregation-based privacy in federated learning using independent component analysis,'' in \emph{International Conference on Machine Learning (ICML)}.\hskip 1em plus 0.5em minus 0.4em\relax PMLR, 2023, pp. 15\,884--15\,899.

\bibitem{gradinversion}
H.~Yin, A.~Mallya, A.~Vahdat, J.~M. Alvarez, J.~Kautz, and P.~Molchanov, ``See through gradients: Image batch recovery via gradinversion,'' in \emph{Proceedings of the IEEE/CVF conference on computer vision and pattern recognition}, 2021, pp. 16\,337--16\,346.

\bibitem{survey_diff_models}
L.~Yang, Z.~Zhang, Y.~Song, S.~Hong, R.~Xu, Y.~Zhao, W.~Zhang, B.~Cui, and M.-H. Yang, ``Diffusion models: A comprehensive survey of methods and applications,'' \emph{ACM Computing Surveys}, vol.~56, no.~4, pp. 1--39, 2023.

\bibitem{sem_metrics}
X.~Sun, N.~Gazagnadou, V.~Sharma, L.~Lyu, H.~Li, and L.~Zheng, ``Privacy assessment on reconstructed images: are existing evaluation metrics faithful to human perception?'' \emph{Advances in Neural Information Processing Systems}, vol.~36, 2024.

\bibitem{dreamsim}
\BIBentryALTinterwordspacing
S.~Fu, N.~Tamir, S.~Sundaram, L.~Chai, R.~Zhang, T.~Dekel, and P.~Isola, ``Dreamsim: Learning new dimensions of human visual similarity using synthetic data,'' in \emph{Advances in Neural Information Processing Systems}, A.~Oh, T.~Naumann, A.~Globerson, K.~Saenko, M.~Hardt, and S.~Levine, Eds., vol.~36.\hskip 1em plus 0.5em minus 0.4em\relax Curran Associates, Inc., 2023, pp. 50\,742--50\,768. [Online]. Available: \url{https://proceedings.neurips.cc/paper_files/paper/2023/file/9f09f316a3eaf59d9ced5ffaefe97e0f-Paper-Conference.pdf}
\BIBentrySTDinterwordspacing

\bibitem{lipsim}
S.~Ghazanfari, A.~Araujo, P.~Krishnamurthy, F.~Khorrami, and S.~Garg, ``Lipsim: A provably robust perceptual similarity metric,'' in \emph{The Twelfth International Conference on Learning Representations (ICLR)}, 2024.

\bibitem{invgrad2}
D.~I. Dimitrov, M.~Balunovic, N.~Konstantinov, and M.~Vechev, ``Data leakage in federated averaging,'' \emph{Transactions on Machine Learning Research}, 2022.

\bibitem{cgir}
X.~Xu, P.~Liu, W.~Wang, H.-L. Ma, B.~Wang, Z.~Han, and Y.~Han, ``Cgir: Conditional generative instance reconstruction attacks against federated learning,'' \emph{IEEE Transactions on Dependable and Secure Computing}, vol.~20, no.~6, pp. 4551--4563, 2023.

\bibitem{zhu2021rgap}
\BIBentryALTinterwordspacing
J.~Zhu and M.~B. Blaschko, ``R-{\{}gap{\}}: Recursive gradient attack on privacy,'' in \emph{International Conference on Learning Representations (ICLR)}, 2021. [Online]. Available: \url{https://openreview.net/forum?id=RSU17UoKfJF}
\BIBentrySTDinterwordspacing

\bibitem{vincent2011}
P.~Vincent, ``A connection between score matching and denoising autoencoders,'' \emph{Neural Computation}, vol.~23, no.~7, pp. 1661--1674, 2011.

\bibitem{imagenet}
J.~Deng, W.~Dong, R.~Socher, L.-J. Li, K.~Li, and L.~Fei-Fei, ``Imagenet: A large-scale hierarchical image database,'' in \emph{2009 IEEE Conference on Computer Vision and Pattern Recognition}, 2009, pp. 248--255.

\bibitem{convir}
Y.~Cui, W.~Ren, X.~Cao, and A.~Knoll, ``Revitalizing convolutional network for image restoration,'' \emph{IEEE Transactions on Pattern Analysis and Machine Intelligence}, pp. 1--16, 2024.

\bibitem{irsde}
Z.~Luo, F.~K. Gustafsson, Z.~Zhao, J.~Sj{\"o}lund, and T.~B. Sch{\"o}n, ``Refusion: Enabling large-size realistic image restoration with latent-space diffusion models,'' in \emph{Proceedings of the IEEE/CVF conference on computer vision and pattern recognition}, 2023, pp. 1680--1691.

\bibitem{irsde2}
\BIBentryALTinterwordspacing
Z.~Luo, F.~K. Gustafsson, Z.~Zhao, J.~Sj\"{o}lund, and T.~B. Sch\"{o}n, ``Image restoration with mean-reverting stochastic differential equations,'' in \emph{Proceedings of the 40th International Conference on Machine Learning}, ser. Proceedings of Machine Learning Research, A.~Krause, E.~Brunskill, K.~Cho, B.~Engelhardt, S.~Sabato, and J.~Scarlett, Eds., vol. 202.\hskip 1em plus 0.5em minus 0.4em\relax PMLR, 23--29 Jul 2023, pp. 23\,045--23\,066. [Online]. Available: \url{https://proceedings.mlr.press/v202/luo23b.html}
\BIBentrySTDinterwordspacing

\bibitem{cao2018vggface2}
Q.~Cao, L.~Shen, W.~Xie, O.~M. Parkhi, and A.~Zisserman, ``Vggface2: A dataset for recognising faces across pose and age,'' in \emph{2018 13th IEEE international conference on automatic face \& gesture recognition (FG 2018)}.\hskip 1em plus 0.5em minus 0.4em\relax IEEE, 2018, pp. 67--74.

\bibitem{survey_diff_neuro}
\BIBentryALTinterwordspacing
J.~Su, B.~Xu, and H.~Yin, ``A survey of deep learning approaches to image restoration,'' \emph{Neurocomputing}, vol. 487, pp. 46--65, 2022. [Online]. Available: \url{https://www.sciencedirect.com/science/article/pii/S0925231222002089}
\BIBentrySTDinterwordspacing

\bibitem{survey_diff_arx}
X.~Li, Y.~Ren, X.~Jin, C.~Lan, X.~Wang, W.~Zeng, X.~Wang, and Z.~Chen, ``Diffusion models for image restoration and enhancement--a comprehensive survey,'' \emph{arXiv preprint arXiv:2308.09388}, 2023.

\bibitem{survey_dp}
N.~Ponomareva, H.~Hazimeh, A.~Kurakin, Z.~Xu, C.~Denison, H.~B. McMahan, S.~Vassilvitskii, S.~Chien, and A.~G. Thakurta, ``How to dp-fy ml: A practical guide to machine learning with differential privacy,'' \emph{Journal of Artificial Intelligence Research}, vol.~77, pp. 1113--1201, 2023.

\bibitem{alistarh2017qsgd}
D.~Alistarh, D.~Grubic, J.~Li, R.~Tomioka, and M.~Vojnovic, ``Qsgd: Communication-efficient sgd via gradient quantization and encoding,'' \emph{Advances in neural information processing systems}, vol.~30, 2017.

\bibitem{aji2017topk}
A.~F. Aji and K.~Heafield, ``Sparse communication for distributed gradient descent,'' in \emph{Proceedings of the 2017 Conference on Empirical Methods in Natural Language Processing}, 2017, pp. 440--445.

\bibitem{vgg16}
K.~Simonyan and A.~Zisserman, ``Very deep convolutional networks for large-scale image recognition,'' \emph{arXiv preprint arXiv:1409.1556}, 2014.

\bibitem{resnet}
K.~He, X.~Zhang, S.~Ren, and J.~Sun, ``Deep residual learning for image recognition,'' in \emph{Proceedings of the IEEE conference on computer vision and pattern recognition}, 2016, pp. 770--778.

\bibitem{lpips}
R.~Zhang, P.~Isola, A.~A. Efros, E.~Shechtman, and O.~Wang, ``The unreasonable effectiveness of deep features as a perceptual metric,'' in \emph{Proceedings of the IEEE conference on computer vision and pattern recognition}, 2018, pp. 586--595.

\bibitem{lion}
X.~Chen, C.~Liang, D.~Huang, E.~Real, K.~Wang, H.~Pham, X.~Dong, T.~Luong, C.-J. Hsieh, Y.~Lu \emph{et~al.}, ``Symbolic discovery of optimization algorithms,'' \emph{Advances in neural information processing systems}, vol.~36, 2024.

\bibitem{adam}
D.~P. Kingma, ``Adam: A method for stochastic optimization,'' \emph{arXiv preprint arXiv:1412.6980}, 2014.

\bibitem{ffhq}
T.~Karras, ``A style-based generator architecture for generative adversarial networks,'' \emph{arXiv preprint arXiv:1812.04948}, 2019.

\bibitem{survey_he}
\BIBentryALTinterwordspacing
A.~Acar, H.~Aksu, A.~S. Uluagac, and M.~Conti, ``A survey on homomorphic encryption schemes: Theory and implementation,'' \emph{ACM Comput. Surv.}, vol.~51, no.~4, Jul. 2018. [Online]. Available: \url{https://doi.org/10.1145/3214303}
\BIBentrySTDinterwordspacing

\bibitem{bonawitz2017practical}
K.~Bonawitz, V.~Ivanov, B.~Kreuter, A.~Marcedone, H.~B. McMahan, S.~Patel, D.~Ramage, A.~Segal, and K.~Seth, ``Practical secure aggregation for privacy-preserving machine learning,'' in \emph{proceedings of the 2017 ACM SIGSAC Conference on Computer and Communications Security}, 2017, pp. 1175--1191.

\bibitem{sok_sa}
M.~Mohamad, M.~Önen, W.~Ben~Jaballah, and M.~Conti, ``Sok: Secure aggregation based on cryptographic schemes for federated learning,'' in \emph{PETS 2023, 23rd Privacy Enhancing Technologies Symposium, 10-15 July 2023, Lausanne, Switzerland (Hybrid Conference)}, IACR, Ed., Lausanne, 2023, iACR.

\bibitem{he_opt}
Q.~Xie, S.~Jiang, L.~Jiang, Y.~Huang, Z.~Zhao, S.~Khan, W.~Dai, Z.~Liu, and K.~Wu, ``Efficiency optimization techniques in privacy-preserving federated learning with homomorphic encryption: A brief survey,'' \emph{IEEE Internet of Things Journal}, vol.~11, no.~14, pp. 24\,569--24\,580, 2024.

\end{thebibliography}
